\documentclass[10pt, journal]{IEEEtran}
\IEEEoverridecommandlockouts
\usepackage{cite}
\usepackage{amsmath,amssymb,amsfonts}
\usepackage{algorithmic}
\usepackage{graphicx}
\usepackage{textcomp}
\usepackage{xcolor}
\usepackage[caption=false,font=normalsize]{subfig}
\usepackage[colorlinks,linkcolor=blue,anchorcolor=blue,citecolor=blue]{hyperref}
\usepackage{balance}
\usepackage{orcidlink}
\usepackage{listings}
\usepackage{booktabs}
\usepackage{makecell}
\usepackage{multirow}
\def\BibTeX{{\rm B\kern-.05em{\sc i\kern-.025em b}\kern-.08em
T\kern-.1667em\lower.7ex\hbox{E}\kern-.125emX}}

\begin{document}

\title{\textsl{EavesDroid}: Eavesdropping User Behaviors via OS Side-Channels on Smartphones
\thanks{Q. Wang, M. Tang, J. Fu are with the
Key Laboratory of Aerospace Information Security and Trusted Computing, Ministry of Education,
School of Cyber Science and Engineering, Wuhan University, Wuhan 430072, China.
Email: wangquancheng@whu.edu.cn, m.tang@126.com, jmfu@whu.edu.cn}
\thanks{\IEEEauthorrefmark{1}Corresponding author: Ming Tang.}
\thanks{}
\thanks{This work has been submitted to the IEEE for possible publication.
Copyright may be transferred without notice,
after which this version may no longer be accessible.}
}

\author{\IEEEauthorblockN{Quancheng~Wang\orcidlink{0000-0002-0313-1853},
Ming~Tang\orcidlink{0000-0003-2218-0164}\IEEEauthorrefmark{1},
Jianming~Fu\orcidlink{0000-0002-4639-5824}}}

\maketitle

\begin{abstract}
As the Internet of Things (IoT) continues to evolve,
smartphones have become essential components of IoT systems.
However, with the increasing amount of personal information stored on smartphones,
user privacy is at risk of being compromised by malicious attackers.
Although malware detection engines are commonly installed on smartphones against these attacks,
attacks that can evade these defenses may still emerge.

In this paper, we analyze the return values of system calls on Android smartphones
and find two never-disclosed vulnerable return values that can leak fine-grained user behaviors.
Based on this observation, we present \textsl{EavesDroid},
an application-embedded side-channel attack on Android smartphones that
allows unprivileged attackers to accurately identify fine-grained user behaviors
(e.g., viewing messages and playing videos) via on-screen operations.
Our attack relies on the correlation between user behaviors and the return values
associated with hardware and system resources.
While this attack is challenging since these return values
are susceptible to fluctuation and misalignment caused by many factors,
we show that attackers can eavesdrop on fine-grained user behaviors
using a CNN-GRU classification model that adopts min-max normalization and multiple return value fusion.
Our experiments on different models and versions of Android smartphones demonstrate that
\textsl{EavesDroid} can achieve 98\% and 86\% inference accuracy for 17 classes of
user behaviors in the test set and real-world settings,
highlighting the risk of our attack on user privacy.
Finally, we recommend effective malware detection, carefully designed obfuscation methods,
or restrictions on reading vulnerable return values to mitigate this attack.
\end{abstract}


\section{Introduction}
\IEEEPARstart{S}{martphones} are essential in the Internet of Things (IoT) systems
and contain large amounts of sensitive data.
In recent years, malicious attacks on these sensitive data have attracted widespread attention.
Although the Android system incorporates several security mechanisms to ensure the confidentiality of user privacy,
there are still vulnerabilities that attackers can exploit to steal sensitive data from smartphones,
such as side-channel attacks~\cite{spreitzer2017systematic} and physical layer attacks~\cite{lu2022reinforcement}.

Side-channel attacks allow attackers to infer sensitive information
by monitoring some affected states or attributes without high-level privileges.
Until now, researchers have discovered a variety of side-channel attacks against Android smartphones.
For example, CPU caches can provide the cache contention state between different applications
and leak sensitive information such as user keystrokes~\cite{wang2019unveiling}.
Then, GPU performance counters~\cite{yang2022eavesdropping} related to graphics rendering
can leak security-critical inputs from users.
Moreover, graphics interrupts~\cite{ma2021effectiveness} can represent unique characteristics
of the GPU workload and provide information about smartphone user behaviors.
In addition, gyroscopes~\cite{gao2022inertiear} and accelerometers~\cite{hu2022accear} interfered with
by speakers playing audio can provide sensitive information such as speech authentication PINs.

Besides CPU~\cite{gruss2019page,gulmezoglu2019undermining,wang2019unveiling,cronin2021exploration},
GPU~\cite{naghibijouybari2018rendered,luo2019side,cronin2021exploration,ma2021effectiveness,yang2022eavesdropping},
and IMU sensors~\cite{javed2020alphalogger,gao2022inertiear,hu2022accear},
many other shared smartphone resources could be potential sources of side-channel attacks
~\cite{spreitzer2018scandroid,spreitzer2018procharvester,zhang2018level,yang2019inference}.
Among them, the OS kernel is considered an effective target for side-channel attacks,
as it provides various system calls for applications to interact with the kernel
and can obtain information about the system state (e.g., memory, filesystem, and network).
ASVAAN~\cite{brussani2022asvaan} is an OS side-channel attack on Android smartphones
that discloses eight vulnerable return values of system calls
and can infer application launching and website browsing with an accuracy of 73\% and 85\%
through the DTW algorithm and the k-fold cross-validation.

However, this attack cannot identify more fine-grained user behaviors,
which are the on-screen operations that users perform on specific applications
by tapping, swiping, scrolling, and other gestures.
These finer-grained user behaviors can also be used for further attacks to cause more privacy leakage.
For example, by inferring the fine-grained user behaviors,
the attacker can obtain the sequence of behaviors of the victim users and
then track the habits and preferences of smartphone users to build their profiles
or de-anonymize users in the network to infer their locations and identities~\cite{ren2018bug}.

To address the limitations of previous work,
it is necessary to investigate whether OS side-channel attacks
can threaten fine-grained user behaviors on smartphones.
Therefore, in this paper, we first analyze the correlation between the vulnerable return values
of system calls found in previous work~\cite{brussani2022asvaan} and fine-grained user behaviors and
find that the return values can provide privacy about fine-grained user behaviors.
We then conduct a comprehensive analysis of all system calls in the Android system
and finally find two new vulnerable return values not considered in the previous work:
\texttt{statvfs.f\_ffree} and \texttt{sysinfo.freeram},
where the former is the most effective return value in inferring fine-grained user behaviors.

\IEEEpubidadjcol

After exploring all the vulnerable return values, we present \textsl{EavesDroid},
an application-embedded OS side-channel attack on smartphones that
enables unprivileged attackers to infer fine-grained user behaviors accurately.
Our attack exploits the correlation between these behaviors and the return values of system calls.
To carry out the attack, We develop a malicious service that can be widely
deployed on different models and versions of Android smartphones.
This service reads vulnerable return values by frequently invoking system calls
in the background when the user performs fine-grained user behaviors.
Finally, the attacker gathers these return values to classify and identify specific user behaviors.

Two significant challenges during the inference process can lead to confusion:
each time series collected is misaligned due to different system states at the time of collection;
other running processes can cause fluctuations in the return values of system calls.
Nevertheless, we demonstrate that \textsl{EavesDroid} can address these challenges
and seriously threaten fine-grained user behaviors with a new CNN-GRU classification model.
Then, we apply min-max normalization to the original time series to
align the time series and thus improve inference accuracy.
In addition, we combine multiple features of the return values to eliminate
the influence of other running processes and further improve inference accuracy.

Afterward, we evaluate \textsl{EavesDroid} on different models and versions of Android smartphones.
The experimental results show that \textsl{EavesDroid} can achieve 98\% and 86\% inference accuracy
for already considered 17 user behaviors in the test set and real-world settings.
Moreover, we also validate that our attack can be launched with over 80\% accuracy
across different device models and application versions.
In addition, the experiments also demonstrate that our attack can
distinguish 41 multi-behavior time sequences with approximately 89\% accuracy
and achieve 90\% inference accuracy on noisy scenarios,
which is sufficient to justify the effectiveness of our attack.
Finally, our attack can also bypass existing static and dynamic anti-malware engines,
further increasing the vulnerability of our attack.

Furthermore, we also discuss some alleviations for the proposed attack.
First, existing detection malware detection methods still leave security holes
for attackers to infer fine-grained user behaviors,
which urges the emergence of effective detection methods.
Second, although obfuscating the vulnerable return values can confuse attackers,
there needs a tradeoff between security and performance/resource overhead.
Third, a more effective and practical approach is to use access control
(such as SELinux) to restrict application privileges from accessing vulnerable return values.
We hope that our proposal will be helpful for future research on countermeasures.

\textbf{Contribution.} To the best of our knowledge, our work is the first that
allows inferring the fine-grained user behaviors via OS side-channel attacks on smartphones.
In summary, we make the following contributions:
\begin{itemize}
\item We prove that the return values of system calls can be used to infer fine-grained user behaviors
and find two new vulnerable return values never disclosed in the previous work after the comprehensive analysis.
\item We propose an application-embedded OS side-channel attack on smartphones called \textsl{EavesDroid},
allowing an unprivileged attacker to invoke system calls and read vulnerable return values
to infer fine-grained user behaviors.
\item We introduce a new CNN-GRU network to classify fine-grained user behaviors.
Then we apply min-max normalization to the original time series and
combine multiple features of return values to improve inference accuracy.
\end{itemize}

\section{Background\label{sec-background}}
In this section, we provide background information and related work
regarding side-channel attacks on smartphones, system calls, and deep learning techniques.

\subsection{Side-Channel Attacks on Smartphones}
Side-channel attacks exploit unintended information leakage from
computing devices or implementations to infer sensitive information~\cite{spreitzer2017systematic}.
And software-based side-channel attacks are the most common attacks on mobile devices
due to the following reasons.
On the one hand, there are numerous shared resources on smartphones,
thus attackers can infer sensitive information from the contention state of shared resources,
such as CPU, GPU, sensors~\cite{sikder2021survey}, Java APIs, and procfs.
On the other hand, attackers can launch their attacks without physically connecting to target devices,
and their attack methods can spread through application stores or networks~\cite{yang2022eavesdropping}.

And researchers have proposed various side-channel attacks on smartphones.
The GPU is a potential side-channel leakage source for launching eavesdropping attacks.
Yang et al.~\cite{yang2022eavesdropping} propose a side-channel attack that
exploits GPU rendering overdraw on smartphones.
By reading and analyzing performance counters related to GPU rendering,
this attack can infer user credentials input with over 80\% accuracy
on various models of Android smartphones.
Graphics interrupts can represent unique features of the GPU workload,
allowing spy processes to infer behaviors of other processes.
Ma et al.~\cite{ma2021effectiveness} evaluate the potential side-channel of graphics interrupts,
the low-level interface between the GPU and the CPU,
and shows that graphics interrupts can leak user behaviors with an average accuracy of 88.2\%.

IMU sensors are another source of side-channel leakage that
can be exploited to compromise user privacy on smartphones.
InertiEAR~\cite{gao2022inertiear} and AccEar~\cite{hu2022accear}
are two recent attacks that exploit the IMU sensors to
perform acoustic eavesdropping attacks on Android smartphones.
InertiEAR uses accelerometers and gyroscopes to
eavesdrop on both the top and bottom speakers of smartphones,
finally achieving a recognition accuracy of 78.8\% and
a cross-device accuracy of 49.8\% across 12 smartphones.
AccEar exploits accelerometer sensors in different scenarios,
including different sample rates, audio volumes, and device models,
to reconstruct any audio played on a smartphone speaker in an unconstrained vocabulary.

In addition, wired and wireless charging are also used to launch side-channel attacks against smartphones.
Charger-Surfing~\cite{cronin2021charger} exploits the correlation between the power line activity
during charging and the dynamic content on the screen to infer the user's on-screen keystrokes
and PIN codes with 92.8\% accuracy.
Moreover, GhostTalk~\cite{wang2022ghosttalk} is another attack that exploits the power line side-channel
by launching an inaudible voice command injection attack on smartphones
that can fool the human ear and multiple live detection models
and can identify the voice of security-critical information
such as passwords and authentication codes with up to 92\% accuracy.
Furthermore, wireless charging is also a potential source of side-channel leakage,
and La et al.~\cite{la2021wireless} propose a side-channel attack that
effectively launches a website fingerprinting attack on Android and iOS smartphones
with over 90\% accuracy by monitoring and analyzing the wireless charger.

\subsection{Linux System Calls and Deep Learning Techniques}
The system calls are the interface between user applications and the Linux kernel.
Linux kernel provides its services to applications through system calls,
and Linux applications interact with the Linux kernel through APIs formed by a combination of system calls.
Moreover, system calls not only help applications perform their functions,
but their return values also reflect the state of the computer system.
For example, \texttt{sysinfo()}~\cite{sysinfo2023linux} returns system statistics,
including the number of running processes, the amount of memory available,
and the time elapsed since the system booted.
Therefore, we can collect time series traces of the system call return values
that reflect the system state and infer some information about the running application.

In addition, deep learning~\cite{menghani2023efficient,dong2021survey}
is a class of machine learning methods that aims to jointly learn data representations
and model parameters from the input dataset.
Among the deep learning architectures, convolutional neural networks (CNN)~\cite{tan2019efficientnet}
are a class of deep and feed-forward neural networks that
perform feature extraction through convolutional and pooling layers.
Since time series data formed by a series of data points indexed in time order exist in many application fields,
it is an essential research topic to classify time series data with deep learning~\cite{ismail2019deep}.
In addition, CNN and its variants have been widely applied to time series classification,
such as 1d-CNN~\cite{tang2020rethinking} and MACNN~\cite{chen2021multi}.
In our study, we read the return values of system calls that reflect the state of the computer system
and convert them into time series traces, then use a customized CNN-GRU network to classify them.

\section{Analyzing Side-Channel Vulnerabilities in Return Values of System Calls\label{sec-analysis}}
In this section, we analyze the return values of system calls on smartphones
and find that return values can leak fine-grained user behaviors.
Moreover, we discover two previously uncovered vulnerable return values that
can disclose fine-grained user behaviors through experiments.

\subsection{Motivation}
Previous research has shown that the return values of system calls can leak
application launch events and website browsing events~\cite{brussani2022asvaan}.
Since identifying fine-grained user behaviors can be used for further attacks to cause more privacy leakage,
it is necessary to investigate whether OS side-channel attacks
can threaten fine-grained user behaviors on smartphones.
On the one hand, the attacker can use fine-grained user behaviors to
track the habits and preferences of smartphone users.
Then, the attacker can use thousands of habits to build user profiles and violate user privacy,
such as precision marketing to sell products to users,
targeted ad delivery, and purposeful intelligence gathering.
On the other hand, the attacker can de-anonymize users in the network with fine-grained user behaviors,
such as inferring the user's location and identity~\cite{ren2018bug}.

Furthermore, we also find that the disclosed vulnerable return values
are related to memory and filesystem resources on smartphones.
Meanwhile, many other return values are also associated with memory, filesystem,
and other resources in the Linux kernel,
such as \texttt{statvfs.f\_ffree} (the number of free blocks).
Therefore, it is interesting to explore whether there are undisclosed vulnerable return values
that can reveal fine-grained user behaviors.

\subsection{Leaking User Behaviors through Return Values}
As mentioned in Section~\ref{sec-background},
the return values of system calls may reflect the state of the system affected by user behaviors.
Hence, we first conduct a preliminary experiment on OPPO K10 using
the Telegram application to verify this correlation.

\begin{figure}[!htbp]
\centering
\begin{lstlisting}
extern "C" JNIEXPORT jint JNICALL Java_com_example_app_DemoService_func(JNIEnv *env, jobject thiz) {
struct sysinfo info{};
sysinfo(&info);
return info.procs;
}
\end{lstlisting}
\caption[]{An example of the malicious native code: reading the number of current processes.}
\label{fig-example}
\end{figure}

\begin{figure}[!htbp]
\centering
\subfloat[different behaviors]{\includegraphics[width=0.5\linewidth]{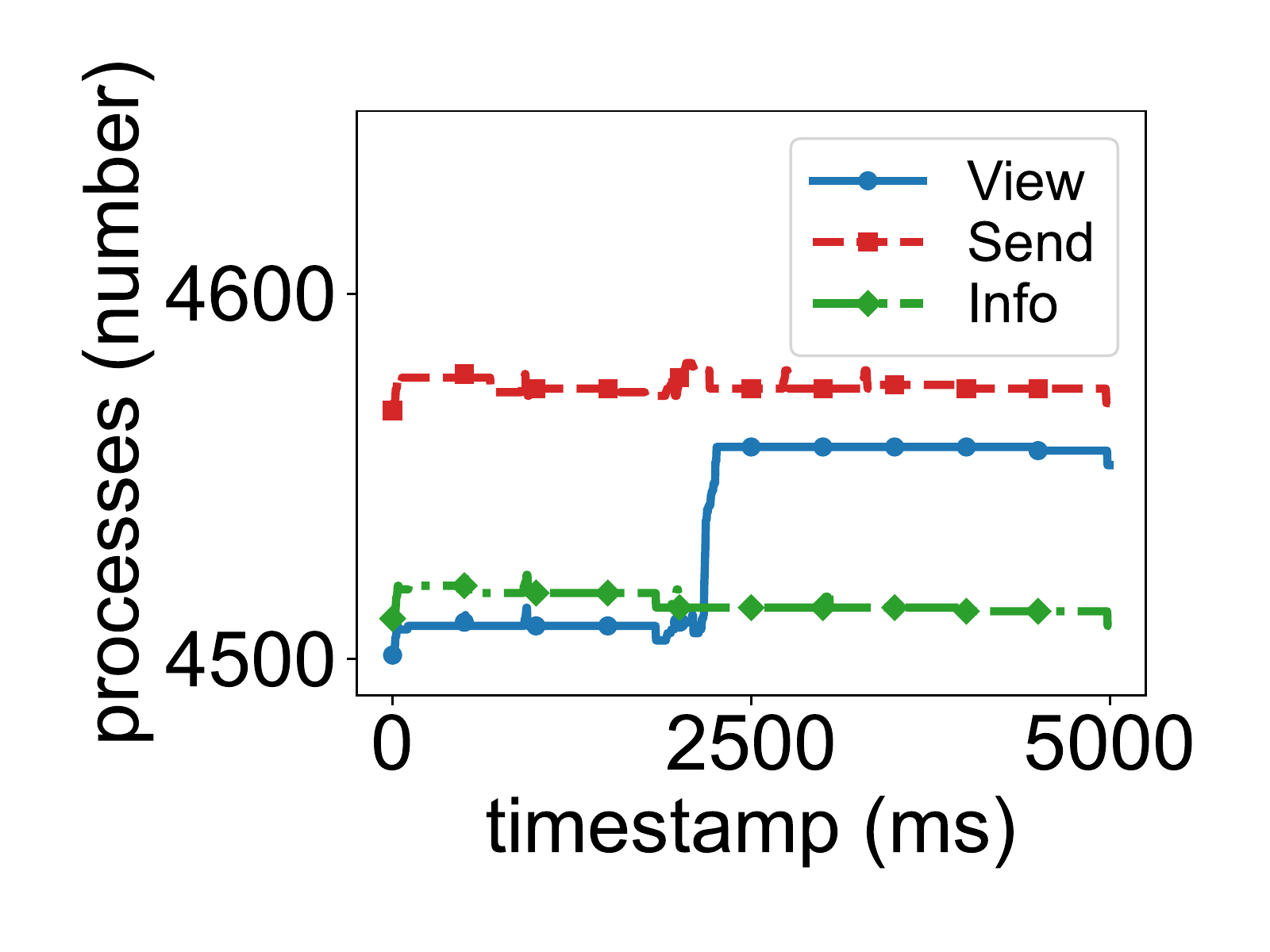}}
\hfill
\subfloat[repetitive behaviors]{\includegraphics[width=0.5\linewidth]{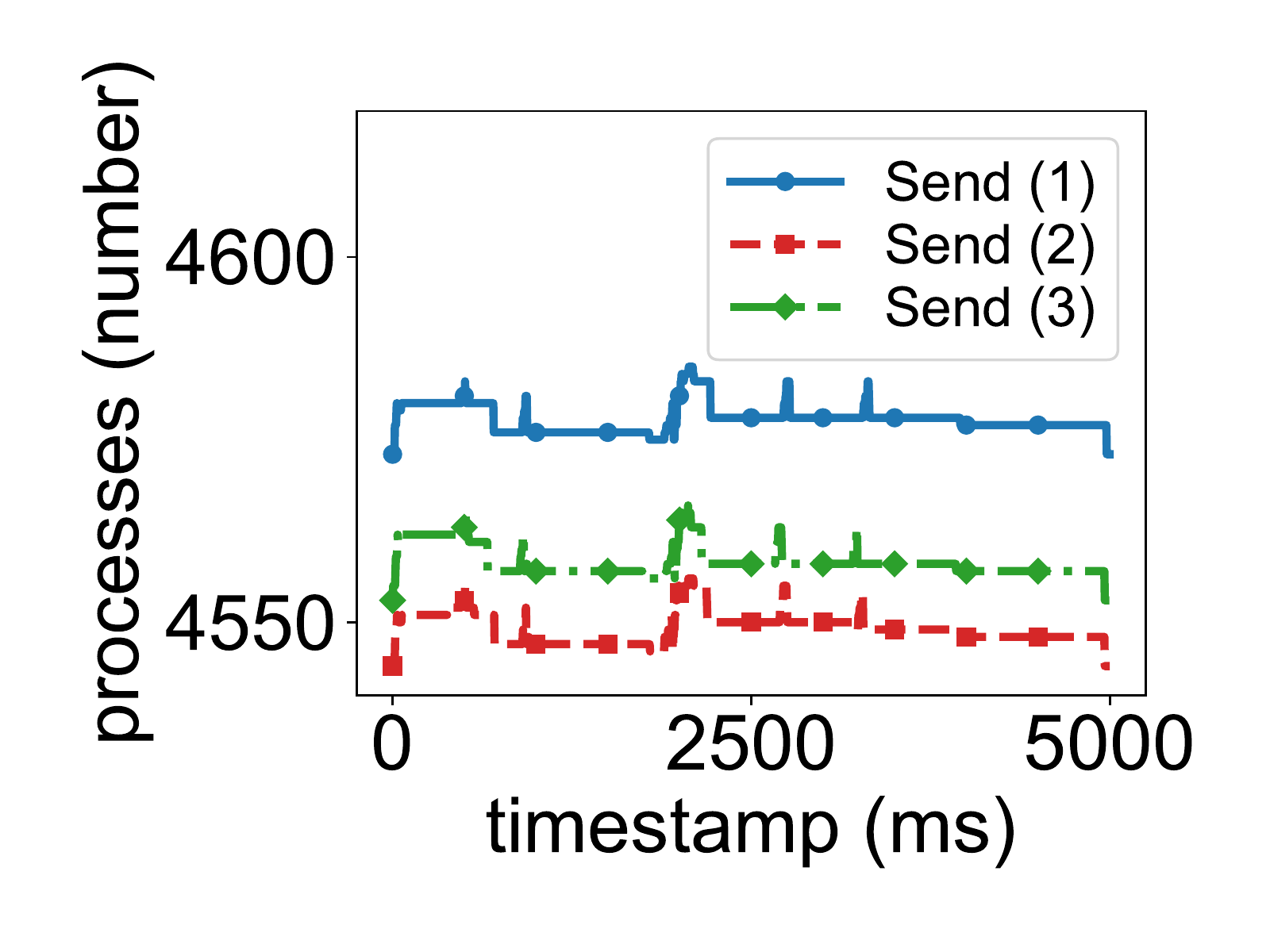}}
\caption[]{The return values of \texttt{sysinfo.procs}
when the user performs different behaviors and repetitive behaviors,
where "Send (1)", "Send (2)" and "Send (3)" denotes different instances of "Send".}
\label{fig-verify}
\end{figure}

Fig.~\ref{fig-example} shows an example of the malicious native code,
where the attacker calls the \texttt{sysinfo()} system call and
repeatedly reads the number of current processes,
then uses this return value to infer fine-grained user behaviors.
In the analysis experiments, malicious native code is embedded in our malicious application
that always runs in the background of the victim's smartphone.
Each time this function executes,
the attacker can read this return value to get the number of running processes.
With a sampling rate of 1000 samples per second and a sampling time of 5 seconds,
the attacker can obtain 5000 data points and use them to generate a time series.
The attacker can then repeat this process to gather multiple time series of
the same user behavior and different user behaviors.
Finally, the attacker tries to classify these time series and infer the user's behavior.

We then run this experiment and collect the return values of \texttt{sysinfo.procs}
when the user performs different and repetitive behaviors.
As shown in Fig.~\ref{fig-verify},
the line shape of return values varies significantly when the user performs different behaviors,
and repetitive user behaviors always result in a similar line shape of return values.
Therefore, we can conclude that the return values of system calls can be used to infer user behaviors.

\subsection{Finding Vulnerable Return Values}
Our primary rationale is that if the user performs the same behavior on the smartphone,
the change in vulnerable return values should be similar.
We calculate the average Euclidean distance between the time series of return values to address their similarity.
Suppose $x$ and $y$ are two different time series data for a return value, and the length of them is $T$,
then the Euclidean distance between $x$ and $y$ is defined as:
\begin{equation}
d(x, y)=\sqrt{\sum_{i=1}^T(x_i-y_i)^2}.
\end{equation}
Assuming the return value is $v$, $N$ is the number of time series data,
and $v[i]$, $v[j]$ represents the $i$-th, $j$-th time series data of $v$,
then we define the average Euclidean distance of $v$ as:
\begin{equation}
avg\_d(v)=\frac{2}{N*(N-1)}\sum_{i=1}^N\sum_{j=i}^N{distance(v[i], v[j])}.
\end{equation}

\begin{table}[!htbp]
\centering
\caption[]{6 Most Vulnerable Return Values on Android Smartphones}
\label{tab-systemcalls}
\begin{tabular}{ccc}
\toprule
System Call & Return Value & Average Distance \\
\midrule
\texttt{statvfs()} & \texttt{f\_ffree} & 3.2806 \\
\texttt{statvfs()} & \texttt{f\_bavail} & 4.3761 \\
\texttt{sysinfo()} & \texttt{freeram} & 6.9541 \\
\texttt{sysinfo()} & \texttt{get\_avphys\_pages} & 6.9546 \\
\texttt{sysconf()} & \texttt{\_SC\_AVPHYS\_PAGES} & 6.9611 \\
\texttt{sysinfo()} & \texttt{procs} & 8.7784 \\
\bottomrule
\end{tabular}
\end{table}

Based on the above principle and the average Euclidean distance formula,
we analyze the return values of system calls provided in the system call table
and repeat launching Telegram 100 times on OPPO K10.
Table~\ref{tab-systemcalls} shows the 6 most vulnerable return values during our analysis.
Apart from the vulnerable return values discovered in previous work,
we find two new vulnerable return values that can be used to infer fine-grained user behaviors:
\begin{itemize}
\item \texttt{statvfs.f\_ffree}: returns the number of free inodes in the filesystem;
\item \texttt{sysinfo.freeram}: returns the amount of available memory in the system.
\end{itemize}
And the average distance of \texttt{statvfs.f\_ffree} is the lowest among all the vulnerable return values.

\begin{figure}[!htbp]
\centering
\subfloat[\texttt{f\_ffree} (different)]{\includegraphics[width=0.5\linewidth]{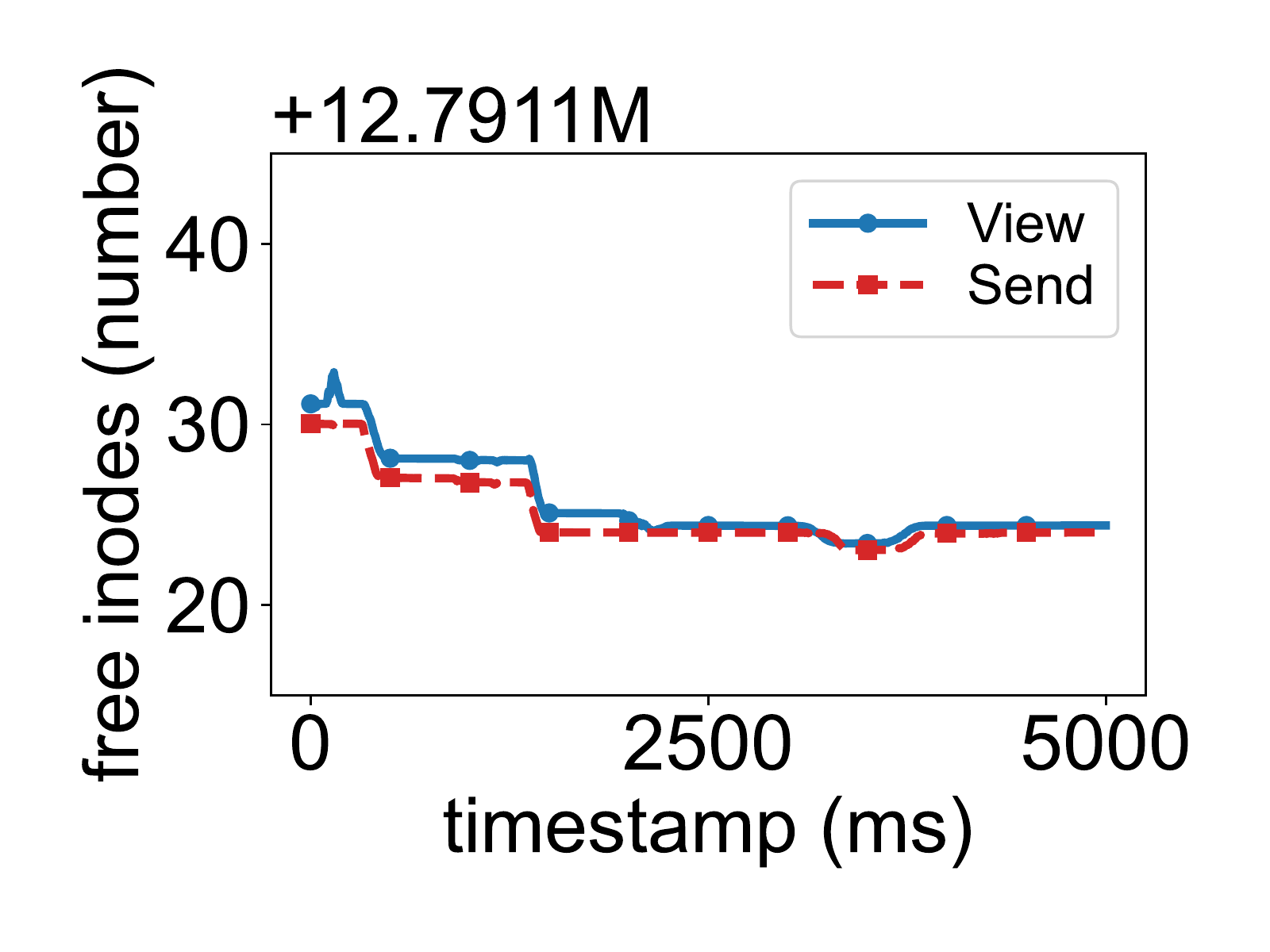}}
\hfill
\subfloat[\texttt{freeram} (different)]{\includegraphics[width=0.5\linewidth]{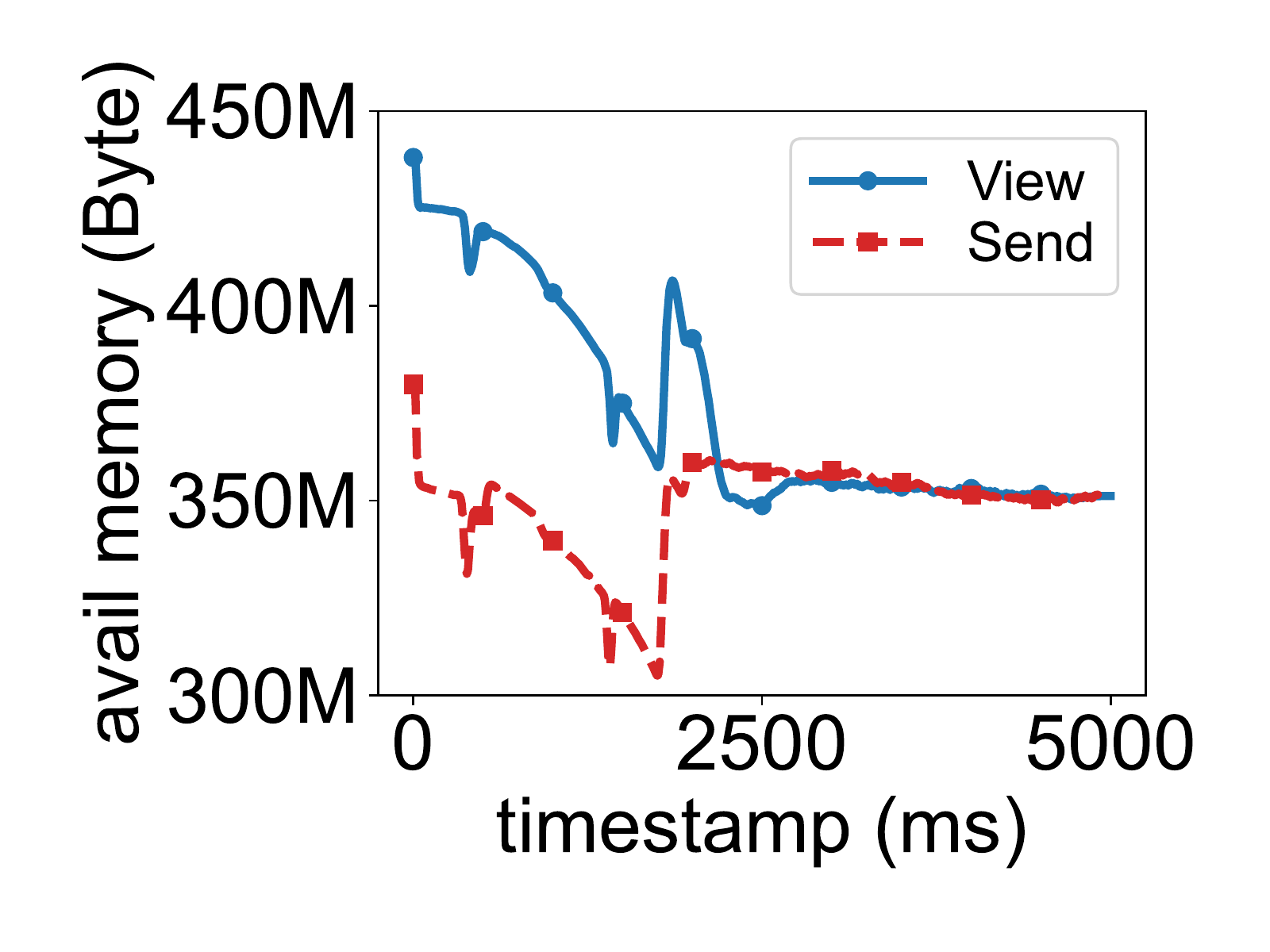}}
\\
\subfloat[\texttt{f\_ffree} (same)]{\includegraphics[width=0.5\linewidth]{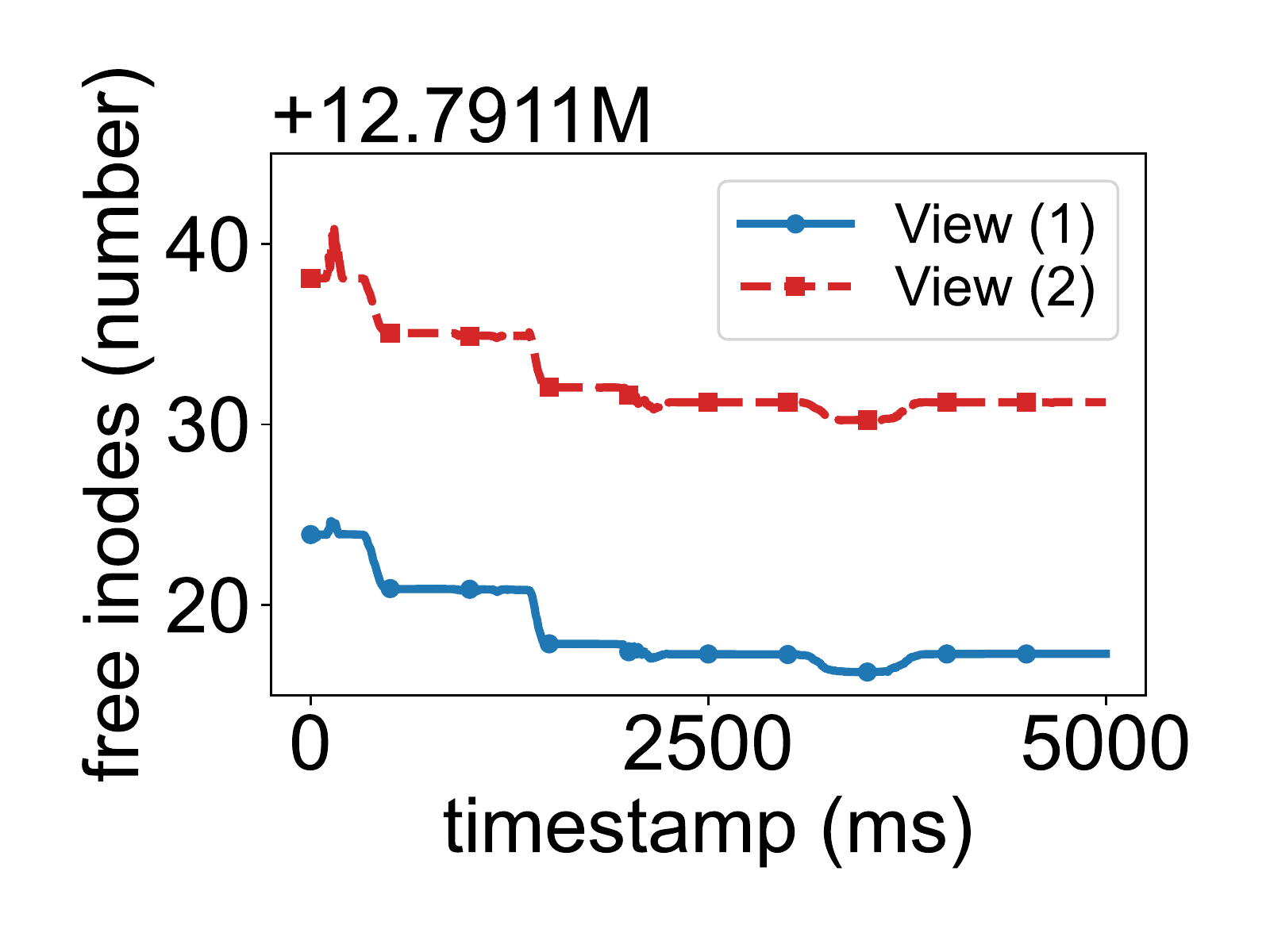}}
\hfill
\subfloat[\texttt{freeram} (same)]{\includegraphics[width=0.5\linewidth]{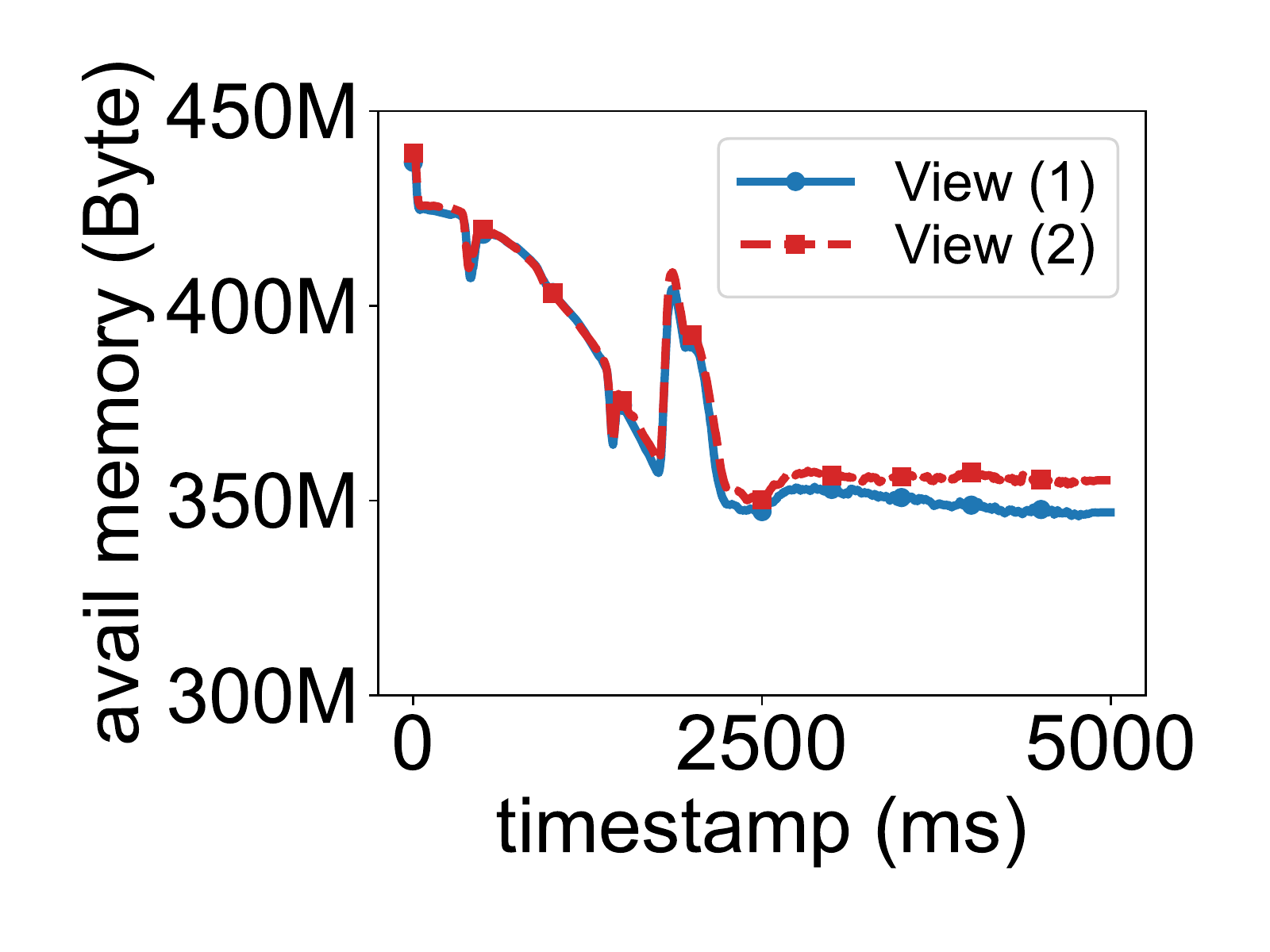}}
\caption[]{Newly disclosed system calls used to infer fine-grained user behaviors:
\texttt{statvfs.f\_ffree} and \texttt{sysinfo.freeram},
where "View (1)" and "View (2)" denote different instances of "View".}
\label{fig-systemcalls}
\end{figure}

As an example shown in Fig.~\ref{fig-systemcalls},
we invoke these two vulnerable return values of system calls.
In Fig.~\ref{fig-systemcalls} (a) and Fig.~\ref{fig-systemcalls} (b),
the blue line represents the average line shape of viewing messages in Telegram,
and the red line represents the average line shape of sending messages in Telegram.
In Fig.~\ref{fig-systemcalls} (c) and Fig.~\ref{fig-systemcalls} (d),
the blue line and the red line both represent the average line shape of viewing messages in Telegram,
where the blue line is the average line shape for the first 50 times,
and the red line is the average line shape for the last 50 times.
The experimental results show that the line shape of the return values
varies significantly for different user behaviors,
and repetitive user behaviors always result in a similar line shape of the return values,
which is consistent with the experimental results in Fig.~\ref{fig-verify} and
demonstrates the effectiveness of using these two newly disclosed return values to infer user behaviors.

\section{EavesDroid Attack}
In this section, we first introduce our threat model
and provide an overview of \textsl{EavesDroid}.
We then describe the details of accessing return values,
gathering device information, and identifying the starting point.
Finally, we discuss the implementation scheme of the attack.

\subsection{Threat Model}
\textbf{Attack Application.}
We assume that our attack application is embedded with malicious code
and can be installed and run on the victim's device,
which is a reasonable assumption for several reasons.
First, the attacker can create a new application that contains malicious code.
Since the malicious code only includes legitimate Linux system calls,
the malicious application can spread through application stores and evade existing malware detection
(see Section~\ref{sec-stealthiness} for more details).
Second, the attacker can modify an existing application, embed malicious code into it,
and then deliver the malicious application to the victim over the Internet.
In addition, the malicious application does not need additional Android permissions
to access the return values of Linux system calls,
and our attack code runs in the background as an Android service.
Therefore, the victim is completely unaware that the malicious application is running.

\textbf{Attack Target.}
First, our attack targets fine-grained user behaviors on their smartphones,
i.e., different user behaviors when using diverse applications,
such as viewing messages and sending messages in social applications (e.g., Telegram, Facebook),
playing videos and refreshing lists in multimedia applications (e.g., YouTube, Instagram).
Second, our attack also targets other information that may violate user privacy,
such as whether a PiP hover window is present or background music is playing
while the user works on another application.

\subsection{Attack Overview}
\begin{figure*}[!htbp]
\centering
\includegraphics[width=\linewidth]{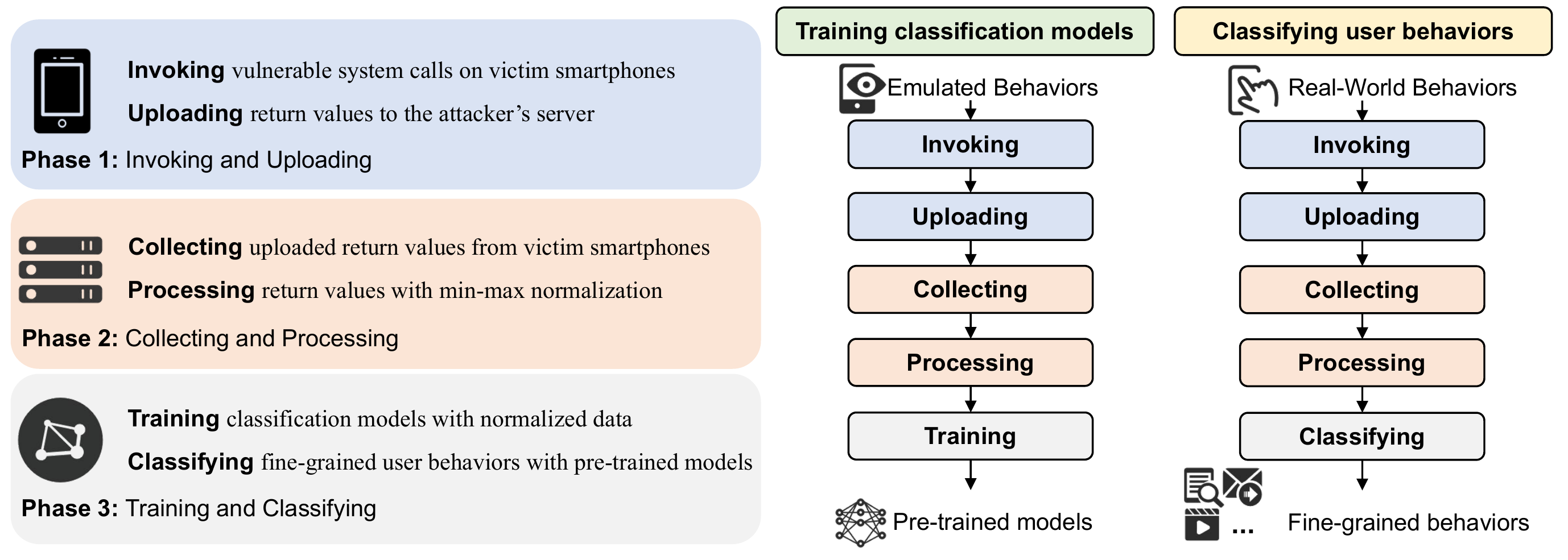}
\caption[]{Overview of \textsl{EavesDroid}: invoking and uploading, collecting and processing, training and classifying.}
\label{fig-overview}
\end{figure*}

As Fig.~\ref{fig-overview} shows, \textsl{EavesDroid} consists of three phases:
invoking and uploading, collecting and processing, and training and classifying.
Before launching the attack on the victim's smartphone,
the attacker emulates user behaviors on devices of different device models
and Android versions to collect sufficient training data.
These data are used to identify correlations between return values and user behaviors,
as well as to build classification models using deep learning techniques.

In the first phase, when the user is using the smartphone,
the attacking application silently reads the return values of system calls and device information
(including model and Android version) from the victim's smartphone in the background,
then uploads the collected data to the attacker's server over the Internet.
In the second phase, the attacker collects data uploaded from the malicious application
over the Internet for further user behavior identification.
Since the raw data collected from smartphones are affected by system states and other factors,
resulting in misalignment and fluctuations,
the attacker preprocesses the data using mix-max normalization to eliminate the effects.
In the third phase, the attacker builds separate classification models for devices
of different models and Android versions with the previously emulated user behaviors.
Then, the attacker classifies and identifies user behavior
using the data normalized in the second phase and the classification model.
In addition, there is no need to retrain the classification model
if the data emulated by the attacker remains the same,
meaning each classification model only needs to be trained once.

\subsection{Accessing Return Values and Necessary Information}
To launch the attack, we need to read the return values of system calls which can reveal user behaviors,
and the device information (model and Android version), which can be used to identify the victim's smartphone.
And then, we need to find a method to covertly transmit the collected data to the attacker,
and keep our malicious application out of the user's sight.

\textbf{Reading.} First, since calling system calls is a pervasive operation
in mobile applications and accessible to unprivileged mobile applications,
we can use the Android NDK to read the return values of system calls on smartphones.
For example, we can get the number of current processes by
reading the \texttt{procs} field of \texttt{sysinfo()},
and obtain the amount of available memory by reading the \texttt{freeram} field of \texttt{sysinfo()}.
Second, to gather information on the device model and Android version for inferring user behaviors,
we read the values of \texttt{android.os.Build} API directly from the victim's smartphone.
We read the values of \texttt{manufacturer}, \texttt{brand}, and \texttt{model} fields
to access the information of device model (e.g., \texttt{"OPPO OPPO PGJM10"} for OPPO K10),
and the values of \texttt{release} and \texttt{sdk\_int} fields to access the information of Android version
(e.g., \texttt{"12 31"} for Android 12).
During our attack, the attacking application periodically invokes the system calls to read return values.
We read the return values at the speed of 1000 calls per second,
and we will further investigate the best reading interval in Section~\ref{sec-impact-interval}.

\textbf{Transmitting.} Since using internal storage (e.g., shared preferences)
or external storage (e.g., SD card) on smartphones requires the user's authorization,
it is not a sound choice to store the collected data on the victim's smartphone
(e.g., \texttt{/sdcard})
and then send the data to the attacker via the Internet or other communication channels.
Nevertheless, the \texttt{INTERNET} permission is granted by default on the Android system.
Whenever the malicious application declares this permission in its manifest file,
it can access the Internet without the user's authorization and upload the collected data to the attacker.
Then the attacker can use the collected data to infer user behaviors.

\textbf{Starting Point.}
Another critical issue is to identify the starting point of user behaviors.
Our preliminary experiment shows that the return value changes drastically
when the user performs a fine-grained behavior (e.g., sending a message),
while the return value tends to be stable when the user does not perform any behavior.
Based on these observations, we can address the starting point of user behaviors by the following two steps:
\begin{itemize}
\item First, we collect the return values of system calls when the user does not perform any behavior,
then use the average change as our baseline.
\item Then, we compare the return values of system calls to the baseline at other times.
If the return value changes significantly from the baseline,
we think the user is performing a specific fine-grained behavior at that time.
\end{itemize}

\subsection{Implementation\label{sec-implementation}}
\textbf{Bot Program.}
For emulating user behaviors, we implement a bot program in Python for automatic data collection.
We simply connect the smartphone to the computer with a USB cable,
then use \texttt{"adb shell"} commands to emulate user behaviors.
For example, we can use the command \texttt{"adb shell input tap 100 100"} to emulate typing on the screen.
And we add random delays between each operation to emulate the varying time interval
between user operations for a realistic emulation.

\textbf{Android Service.}
For the first phase, we implement an Android service in Kotlin to collect the return values of system calls
and register the service by modifying the \texttt{AndroidManifest.xml} file.
In addition, we invoke system calls and read return values using the Android NDK,
a native development kit that allows C and C++ code to be used in Android applications.
We also use \texttt{okhttp3}, a popular third-party web library for Android,
to upload the collected data to the server using the HTTP protocol.

\textbf{Django.}
For the second phase, we implement a web application using Django,
an open-source web framework written in Python.
We use this web application to collect the data uploaded by the malicious application,
and then use the data preprocessing algorithms to eliminate fluctuations and misalignment.
We also embed classification models to classify user behaviors into this web application.

\textbf{Keras.}
For the third phase, we build separate classification models for different devices
using Keras, an open-source neural network library written in Python.
It allows rapid implementation of deep neural networks and
provides user-friendly, modular, and scalable benefits.

\section{A New CNN-GRU Network for Fine-Grained User Behavior Classification}
In this section, we introduce a new CNN-GRU network for
fine-grained user behavior classification.
We then use min-max normalization and multiple feature combinations
to eliminate data misalignment and improve inference accuracy.

\subsection{The Architecture of the CNN-GRU Network}
We use convolutional neural networks (CNN) and gate recurrent neural networks (GRU)
to extract features of user behaviors and build classification models.
And we also customize the structure and parameters of the CNN-GRU network
according to the characteristics of our time series related to user behaviors.
Table~\ref{tab-cnn-gru} summarizes the architecture of the CNN-GRU network,
which includes convolutional layers, pooling layers, a GRU layer,
a fully connected layer, and an output layer.

There are several reasons why we choose CNN and GRU networks as the basic building blocks of our model:
\begin{itemize}
\item First, CNN is the state-of-art solution for
extracting features from images and videos~\cite{xie2017genetic}.
And the data format of our time series representing user behavior is similar to a two-dimensional image,
where the x-axis represents the time point while the Y-axis represents the return value of some system call.
Therefore, we use CNN to extract features from the time series related to user behaviors.
\item Second, GRU is a variant of recurrent neural networks (RNN) for modeling sequential data.
The time series associated with user behavior is also a kind of sequential data,
where the order and the correlation of the data between each time point are of great significance.
Meanwhile, compared with LSTM, GRU has fewer parameters, converges faster,
and performs better than LSTM with accuracy guaranteed~\cite{fu2016using}.
Therefore, we choose GRU as the basic building block of our model
instead of using only CNN.
\item Finally, we also consider the computational complexity of the model.
If we only use GRU to model and classify user behaviors,
the model will be too complex to result in a slow training process.
Therefore, we combine the advantages of CNN and GRU to build the CNN-GRU classification model,
while the former can extract features and reduce data dimensionality,
and the latter can model sequential data.
We will also evaluate the impact of the CNN-GRU network in Section~\ref{sec-impact-cnn-gru}.
\end{itemize}

\begin{table}[!htbp]
\centering
\caption[]{The CNN-GRU Architecture for Classifying User Behaviors}
\label{tab-cnn-gru}
\begin{tabular}{cccc}
\toprule
Layer & Input & Parameters & Output \\
\midrule
Reshape & (5000, n) & Reshape(10) & (500, n * 10) \\
\midrule
Conv\_1 & (500, n * 10) & \makecell[c]{Conv(64, 3, leaky\_relu) \\ MaxPooling(2) \\ BN} & (249, 64) \\
\midrule
Conv\_2 & (249, 64) & \makecell[c]{Conv(128, 3, leaky\_relu) \\ MaxPooling(2) \\ BN} & (123, 128) \\
\midrule
Conv\_3 & (123, 128) & \makecell[c]{Conv(256, 3, leaky\_relu) \\ MaxPooling(2) \\ BN} & (60, 256) \\
\midrule
GRU & (60, 256) & \makecell[c]{GRU(128) \\ BN} & (60, 128) \\
\midrule
FC & (60, 128) & \makecell[c]{FC}  & (7680) \\
\midrule
Output & (7680) & \makecell[c]{Dense(c, softmax)} & (c) \\
\bottomrule
\end{tabular}
\end{table}

In the architecture, the input size is $5000*n$, where $n$ is the number of system calls combined and
$5000$ is the number of timestamps for each fine-grained user behavior.
The input features are first reshaped and extracted with three convolutional layers.
In these convolutional layers:
\begin{itemize}
\item The number of filters is $64$, $128$, and $256$ for the three convolutional layers.
\item The kernel size is $2$, the stride is $1$,
and the padding is the same for each convolutional layer.
\item The activation function is $LeakyReLU$,
and the leaky ratio is $0.3$ for each convolutional layer.
\item A max pooling layer with the kernel size $3$ and
a batch normalization layer are followed by each convolutional layer.
\end{itemize}

A GRU layer with $128$ hidden nodes is then
applied to aggregate the features and followed by a fully connected layer.
The output of the fully connecting layer is connected to the output layer
with the $softmax$ activation function,
where the number of nodes $c$ is equal to the number of user behaviors.
The network is trained with an ADAM optimizer with a learning rate of $0.001$,
a sparse categorical cross-entropy loss function, $100$ epochs, and a mini-batch size of $32$.

During the training process, we first label the emulated data from victim smartphones.
Then, we divide the labeled data into training and test sets,
which are used to train the network and classify the user behaviors, respectively.
The input of the CNN-GRU network is the data preprocessed by the min-max normalization algorithm.

\subsection{Min-Max Normalization for Eliminating Misalignment}
The raw data collected from the smartphones are affected by different system states,
leading to data misalignment, which poses a problem for accurate classification.

To solve this problem, we normalize the data using min-max normalization,
which is one of the most popular data normalization methods.
For each feature of the data, we find the minimum and maximum values of the feature,
then set the minimum value to $0$ and the maximum value to $1$,
while every other value is normalized to a decimal between $0$ and $1$.
Consequently, we apply the min-max normalization to the raw time series data
to eliminate the misalignment and linearize the data in the range of [$0$, $1$]:
\begin{equation}
\begin{split}
x^\ast_{i,j} = \frac{x_{i,j} - \min\limits_{i=1}^{N} x_{i,j}}
{\max\limits_{i=1}^{N} x_{i,j} - \min\limits_{i=1}^{N} x_{i,j}},
\end{split}
\end{equation}
where $x_{i,j}$ is the value of the $i$-th time point of the $j$-th feature,
$N$ is the number of time points, and $x^\ast_{i,j}$ is the normalized value of $x_{i,j}$.

\begin{figure}[!htbp]
\centering
\subfloat[\texttt{procs} (raw)]{\includegraphics[width=0.5\linewidth]{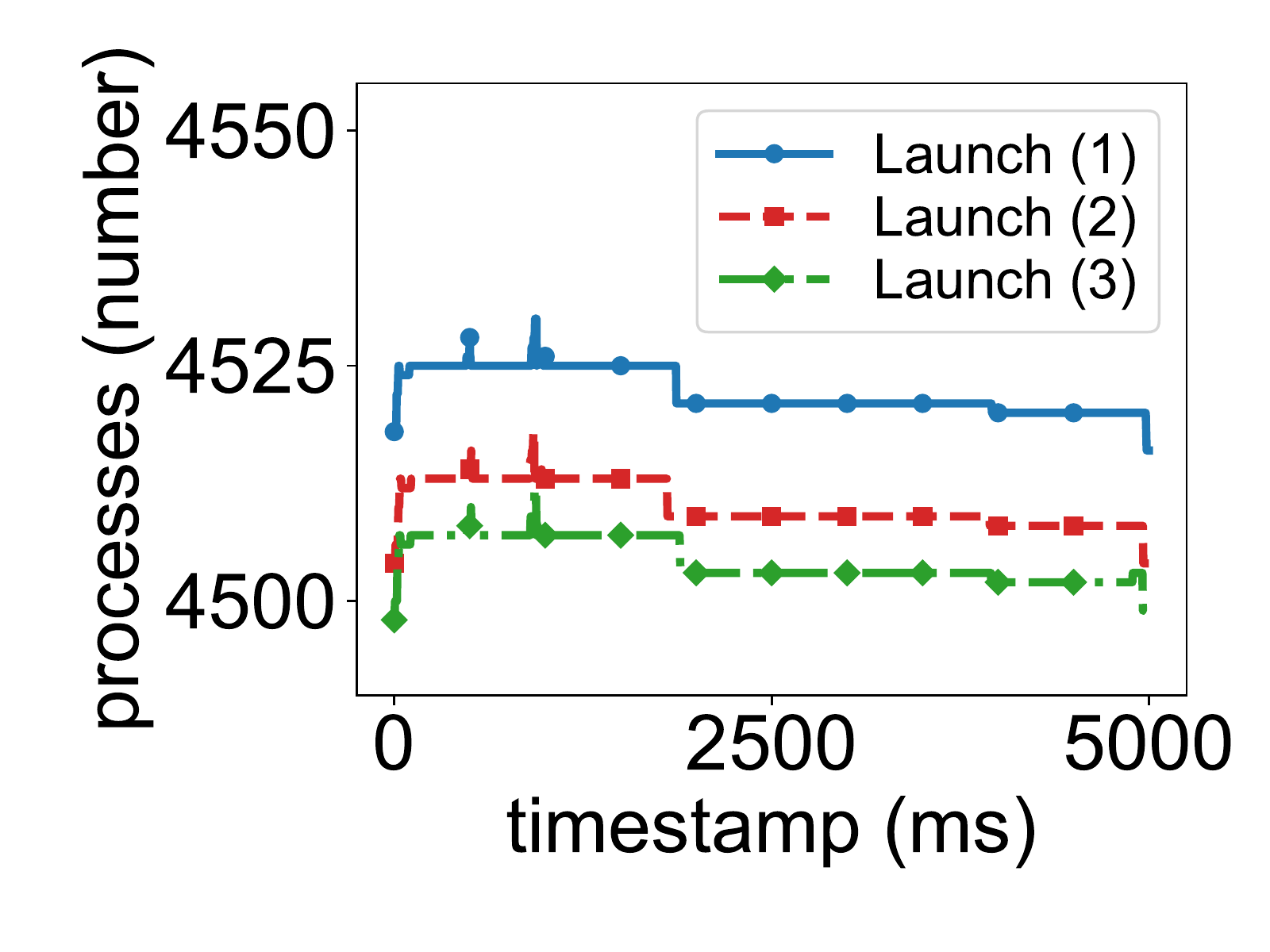}}
\hfill
\subfloat[\texttt{procs} (normalized)]{\includegraphics[width=0.5\linewidth]{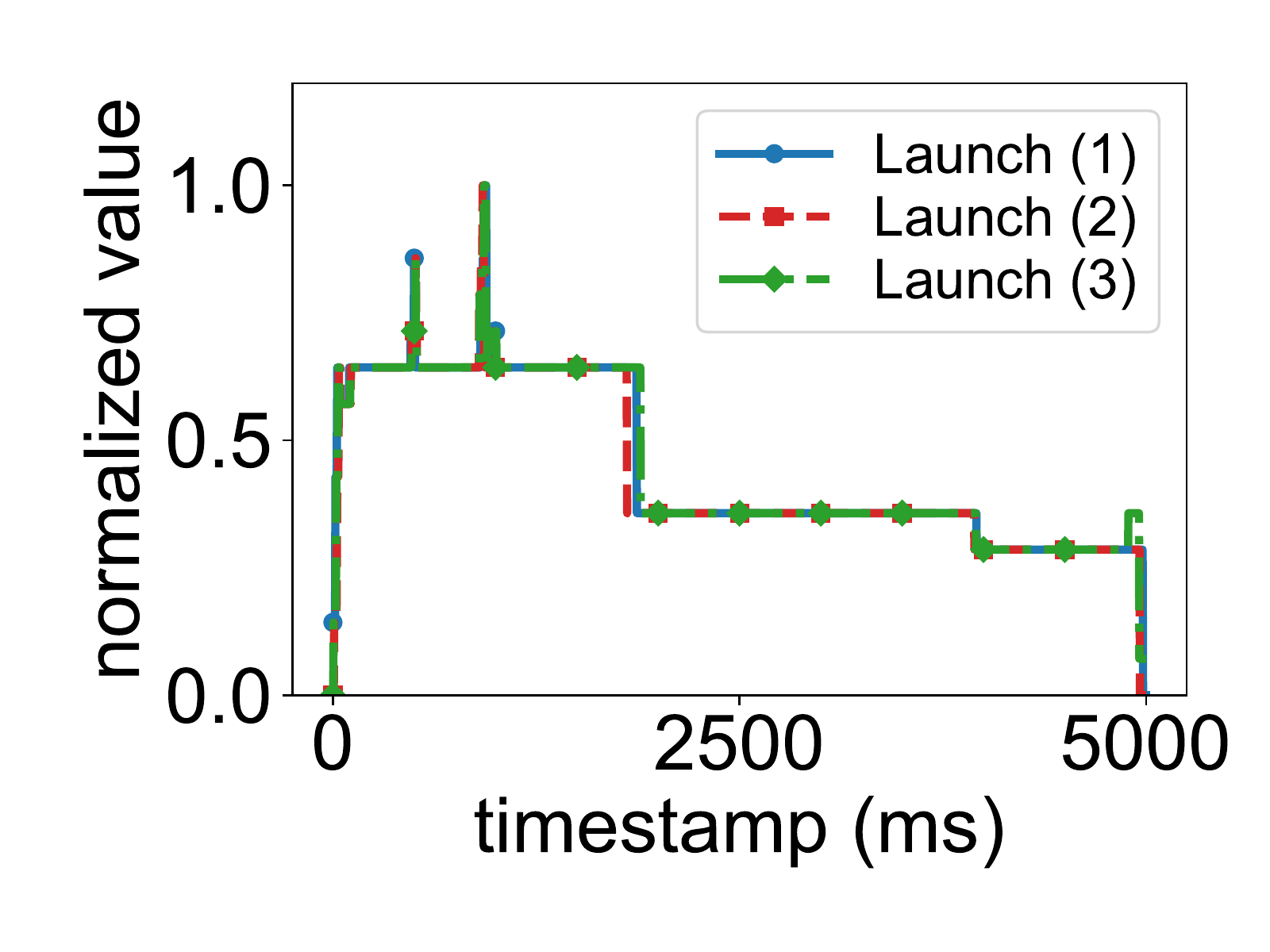}}
\caption[]{The difference between the raw data and the normalized data of \texttt{sysinfo.procs} for launching Telegram,
where "Launch (1)", "Launch (2)" and "Launch (3)" denotes different instances of "Launch".}
\label{fig-normalization}
\end{figure}

Fig.~\ref{fig-normalization} (a) and Fig.~\ref{fig-normalization} (b) show the raw data and the normalized data
of \texttt{sysinfo.procs} for launching Telegram respectively.
Although each time series in the raw data shown in Fig.~\ref{fig-normalization} (a) has similar line shapes,
there is still a misalignment that can affect the classification accuracy.
Then the misalignment is eliminated after mix-max normalization
as shown in Fig.~\ref{fig-normalization} (b).
And we will further investigate the impact of the normalization
on the inference accuracy in Section~\ref{sec-impact-minmax}.

\subsection{Multi-Feature Combination for Improving Accuracy}
We also realize that there are some different user behaviors,
where the line shape of one return value is similar while
the line shape of another return value is quite different.

\begin{figure}[!htbp]
\centering
\subfloat[\texttt{procs} indistinguishable]{\includegraphics[width=0.5\linewidth]{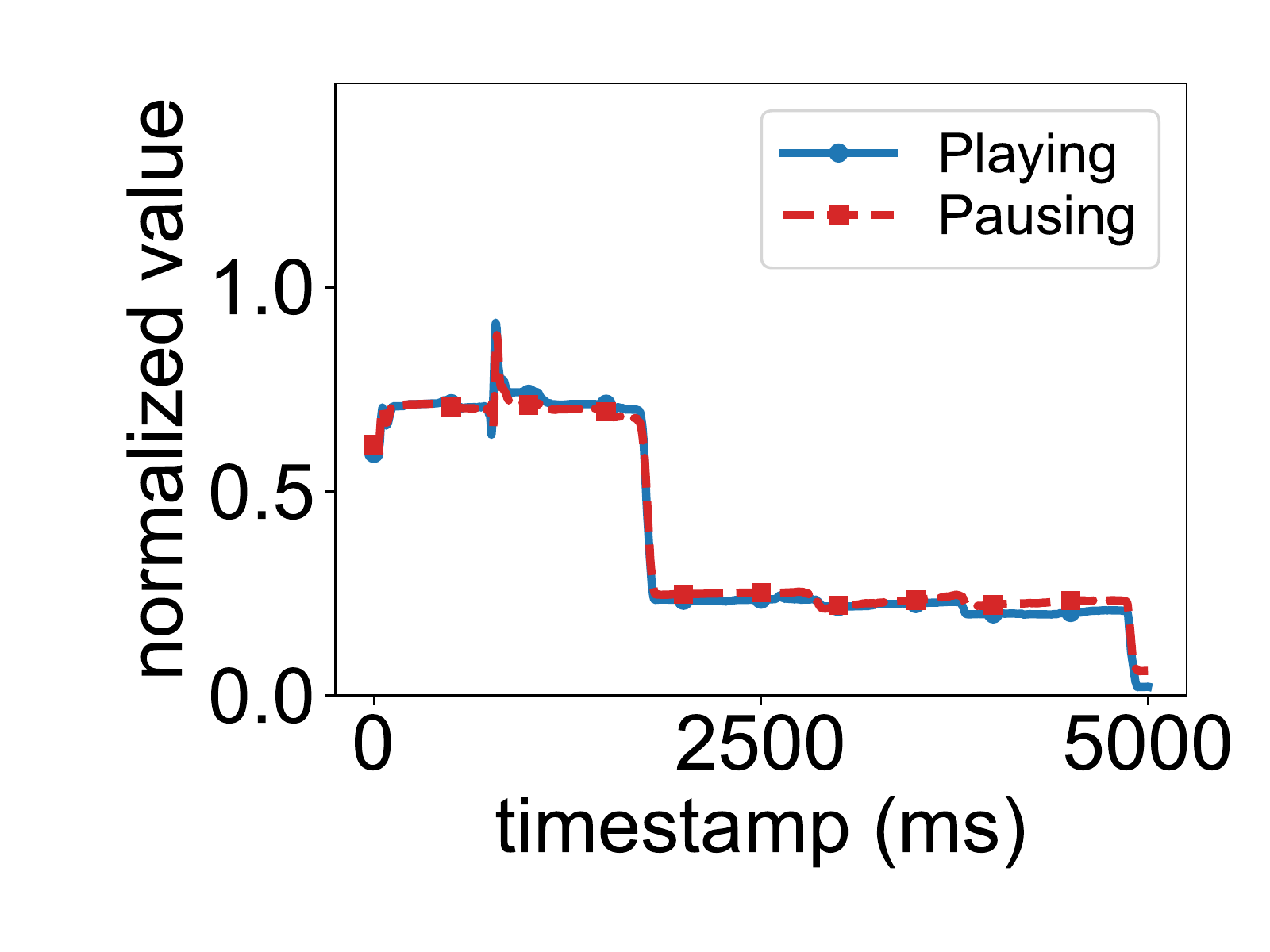}}
\hfill
\subfloat[\texttt{f\_bavail} distinguishable]{\includegraphics[width=0.5\linewidth]{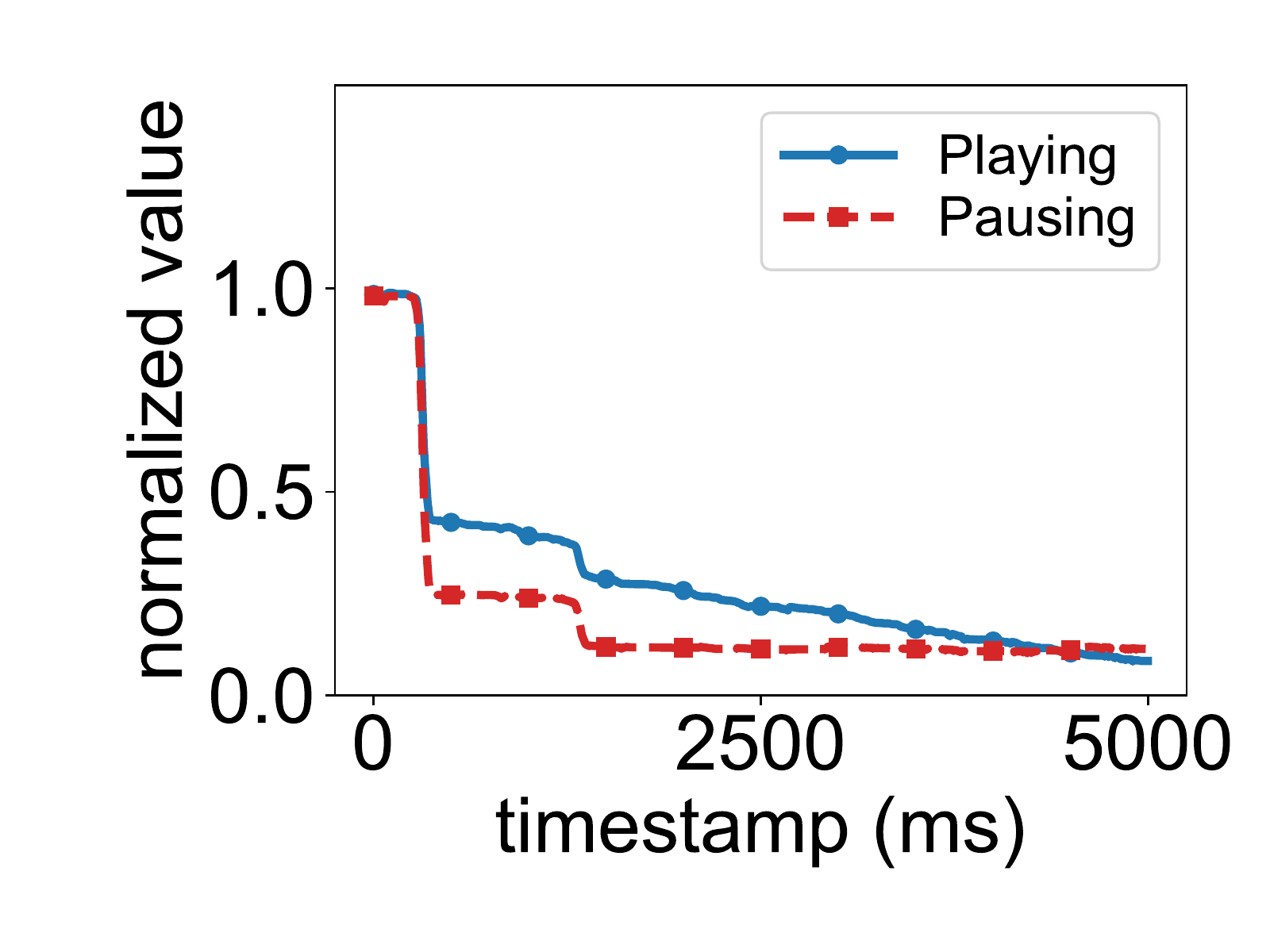}}
\\
\subfloat[\texttt{procs} distinguishable]{\includegraphics[width=0.5\linewidth]{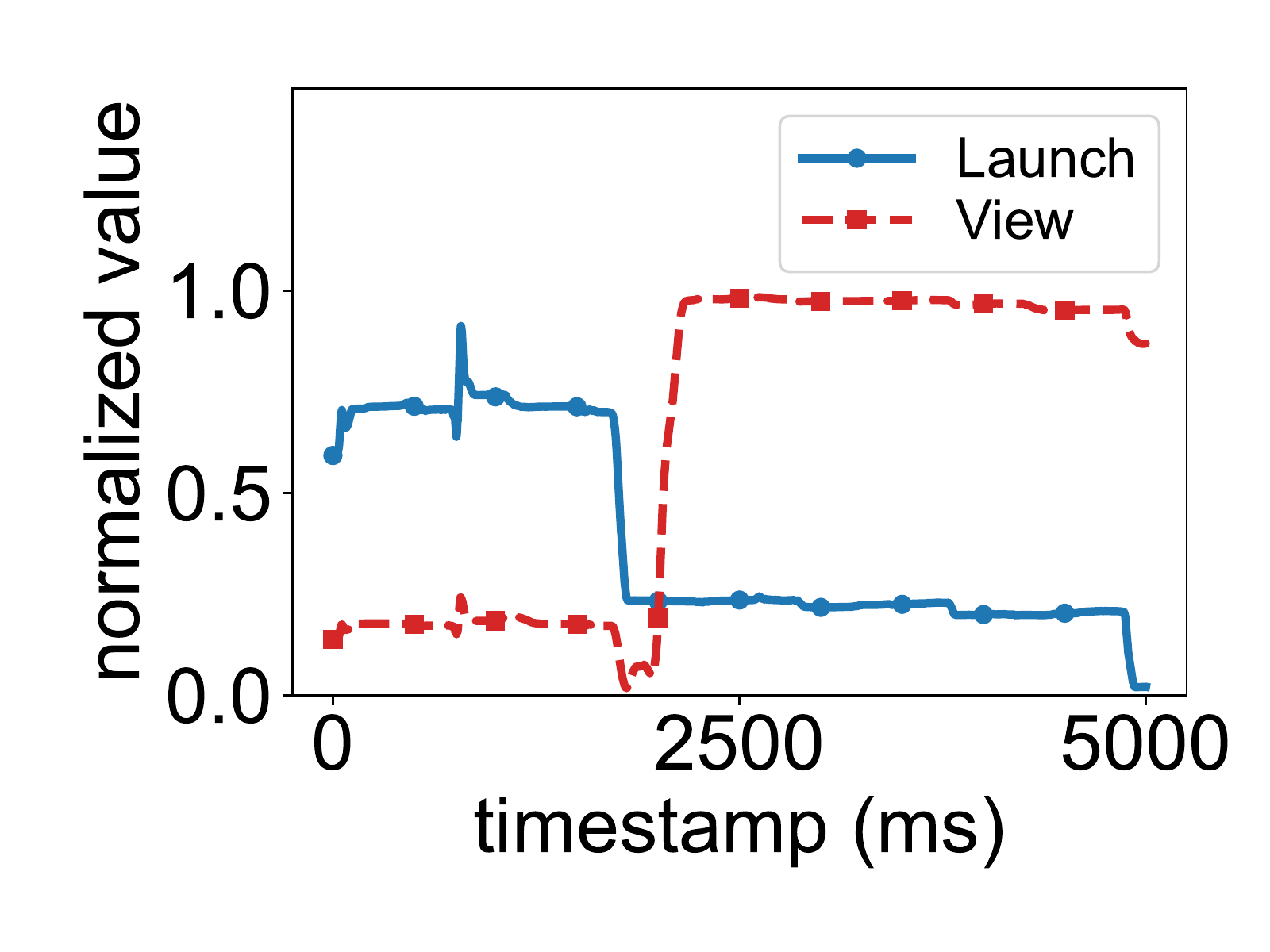}}
\hfill
\subfloat[\texttt{f\_bavail} indistinguishable]{\includegraphics[width=0.5\linewidth]{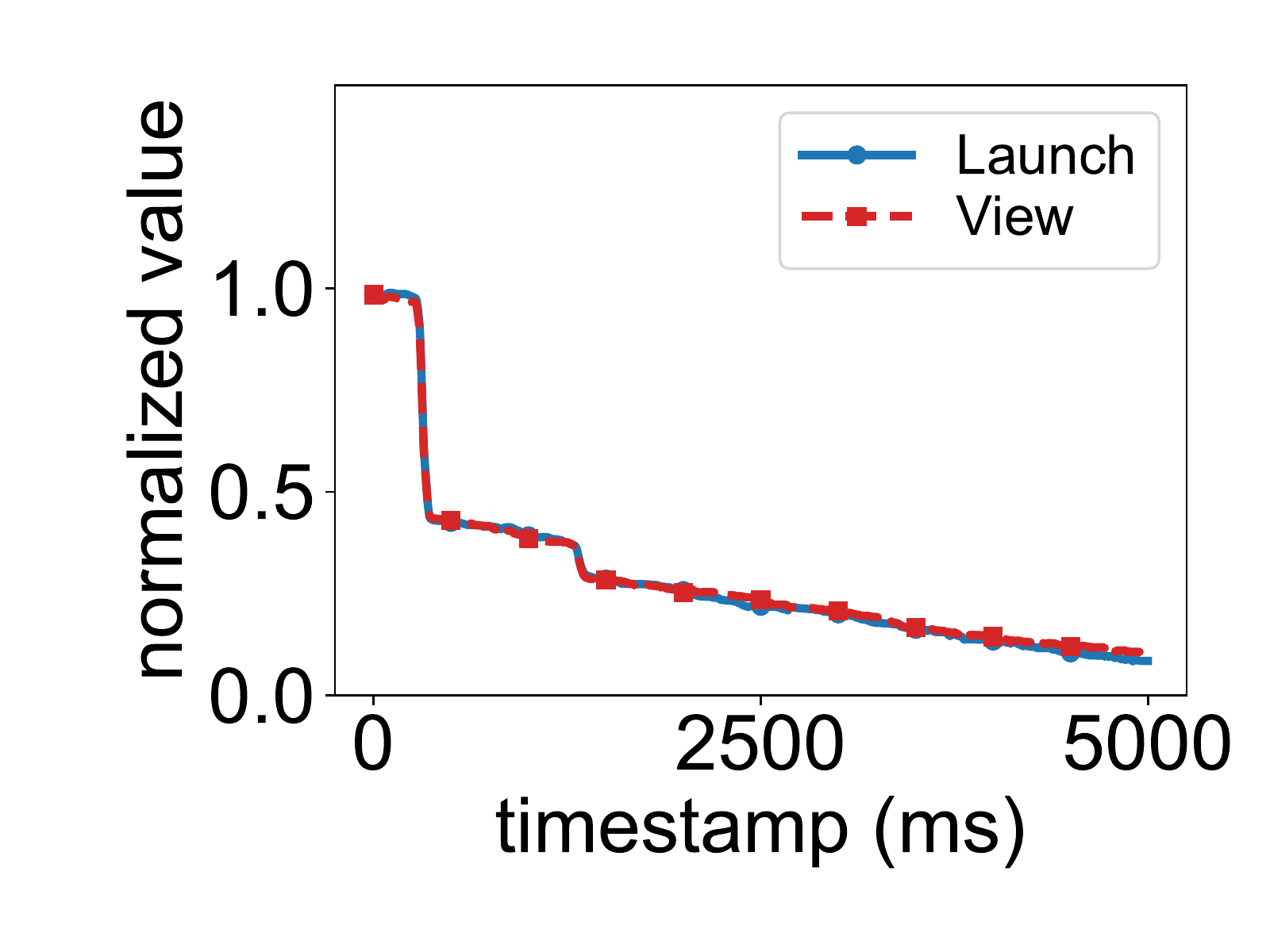}}
\caption[]{The line shape of user behaviors cannot be distinguished by only one feature
(\texttt{sysinfo.procs} or \texttt{statvfs.f\_bavail})
but can be distinguished by combining these two features.}
\label{fig-mul-combine}
\end{figure}

Let us consider the following two pairs of user behaviors:
\begin{itemize}
\item \textit{Launching while a PiP hover window is playing}
and \textit{Launching while a PiP hover window is pausing};
\item \textit{Launching the Telegram application}
and \textit{Viewing messages in the Telegram application}.
\end{itemize}
First, for the first pair of user behaviors,
as shown in Fig.~\ref{fig-mul-combine} (a) and Fig.~\ref{fig-mul-combine} (b),
the blue and red lines represent launching Telegram while a PiP hover window is playing and pausing.
We cannot distinguish them with \texttt{sysinfo.procs} but with \texttt{statvfs.f\_bavail}.
Second, for the second pair of user behaviors,
as shown in Fig.~\ref{fig-mul-combine} (c) and Fig.~\ref{fig-mul-combine} (d),
the blue and red lines represent launching Telegram and viewing messages in Telegram separately.
In this case, we can distinguish them with \texttt{sysinfo.procs}
but cannot distinguish them with \texttt{statvfs.f\_bavail}.
From the above experiments, we can conclude that we cannot fully identify these user behaviors
with either \texttt{sysinfo.procs} or \texttt{statvfs.f\_bavail}.
But if we combine these two features, the problem is solved.

Therefore, to identify more fine-grained user behaviors and improve classification accuracy,
we combine $n$ different features, so the input size is $n*t$,
where $n$ is the number of combined system calls, and $t$ is the number of time points.
And we will further evaluate the impact of feature dimensions
on the accuracy in Section~\ref{sec-impact-feature}.

\section{Evaluation of EavesDroid Attack and CNN-GRU Classification Model}
We present a thorough analysis and evaluation of our proposed \textsl{EavesDroid} attack in this section.
First, we describe the experimental setup and evaluate the classification accuracy
of 17 categories of user behaviors in a laboratory setting.
Second, we have mentioned in previous sections that
our customized CNN-GRU neural network contributes to improving inference accuracy.
Therefore, we compare the CNN-GRU neural network
with the traditional DTW-KNN method and other types of neural networks,
compare the different data preprocessing methods,
and compare the accuracy with the different number of features
to demonstrate the effectiveness of our scheme.
Third, since our malicious service needs to continuously read system call return values
on the victim's smartphone and interact with the attacker's server,
it is critical to evaluate how much overhead our attack application imposes on the victim's smartphone,
including power consumption, CPU utilization, memory usage, and network traffic,
and whether the classification model can identify user behaviors promptly.
So far, all of our evaluations have been conducted on the same smartphone
against the identical version of the applications.
However, the user may use different devices and different versions of applications,
which motivates us to evaluate the adaptability of our attack in such scenarios.
In addition, user behaviors in real-world scenarios are not invariant,
such as the user may perform multiple operations in the same period.
Therefore, we consider three complex cases and evaluate the robustness of such complicated situations.
After our systematic and complete analysis in the laboratory setting,
we also need to evaluate the inference accuracy for user behaviors in real-world conditions
and whether our malicious code can evade the detection by existing anti-malware suites.

\subsection{Experimental Setup\label{sec-setup}}
\textbf{Devices.}
Throughout this work, we use the following Android smartphones:
\begin{itemize}
\item OPPO K10 with Dimensity 8000-MAX and Android 12.
\item Redmi K50 with Dimensity 8100 and Android 12.
\item Xiaomi 9 with Snapdragon 855 and Android 11.
\item Xiaomi 9 with Snapdragon 855 and Android 12.
\item OnePlus 7Pro with Snapdragon 855 and Android 12.
\end{itemize}

\textbf{System Calls.}
We have already analyzed the most vulnerable return values on Android smartphones in Section~\ref{sec-analysis},
and we choose 5 of them to infer fine-grained use behaviors,
where the first three are disclosed in previous work~\cite{brussani2022asvaan},
while the last two are discovered in our study:
\begin{itemize}
\item \texttt{sysinfo.procs}: Number of current processes.
\item \texttt{statvfs.f\_bavail}: Number of free blocks for unprivileged users.
\item \texttt{sysconf.\_SC\_AVPHYS\_PAGES}: Number of currently available pages of physical memory.
\item \texttt{statvfs.f\_ffree}: Number of free inodes.
\item \texttt{sysinfo.freeram}: Size of available memory.
\end{itemize}
By default, the return values of the selected system calls
are read every 1 ms and collected for 5 seconds.

\textbf{Applications.}
Several widely used applications are our target applications: Telegram, YouTube, Gmail, and OneNote.
And for each application, we choose a set of user behaviors (e.g., sending messages, sending emails)
that are more sensitive from the user's perspective to identify.
The list of these applications and user behaviors is shown in Table~\ref{tab-applications}.

\begin{table}[!htbp]
\setlength\tabcolsep{3pt}
\scriptsize
\centering
\caption[]{Details of Applications and Fine-Grained User Behaviors in Our Experiments and the Accuracy of Classification}
\label{tab-applications}
\begin{tabular}{cccc}
\toprule
Application & User Behavior & Description & Accuracy \\
\midrule
Telegram & \makecell[c]{Launch App \\ View Messages \\ Send Messages \\ View Profile} &
\makecell[c]{Launch the Telegram application. \\ Open a chat page from the chat list. \\
Send a message to others in a chat. \\ Open the user profile page from the menu.}
& \makecell[c]{0.9855}\\
\midrule
YouTube & \makecell[c]{Launch App \\ Refresh Videos \\ View Videos \\ Short Videos \\ Search Videos} & 
\makecell[c]{Launch the YouTube application. \\ Refresh the video list on the home page. \\
Open a video from the video list. \\ Open a short video by tapping the button. \\ Search videos with keywords.}
& \makecell[c]{0.9684} \\
\midrule
Gmail & \makecell[c]{Launch App \\ View Emails \\ Send Emails \\ Search Emails} &
\makecell[c]{Launch the Gmail application. \\ Open an email from the email list. \\
Email a friend or person. \\ Search emails with keywords.}
& \makecell[c]{1.0000} \\
\midrule
OneNote & \makecell[c]{Launch App \\ View Notes \\ Create Notes \\ Search Notes} &
\makecell[c]{Launch the OneNote application. \\ Open a note from the note list. \\
Create a new note and save it. \\ Search notes with keywords.}
& \makecell[c]{0.9834} \\
\midrule
All apps & All behaviors & All applications and all user behaviors.
& \makecell[c]{0.9809} \\
\bottomrule
\end{tabular}
\end{table}

\textbf{Additional Behaviors.}
Besides, we also consider some other user behaviors not included in the above list.
Our basic principle is that while the user is operating on the foreground application,
there is a high probability that the video or music is playing in the background.
Our goal is to identify these two types of behaviors:
\begin{itemize}
\item State of the background music (playing).
\item State of the PiP hover window (playing, pausing).
\end{itemize}

\textbf{Dataset.}
In our dataset, each user behavior is emulated 200 times
separately on each experimental device.
We first define the user behaviors to be analyzed in our experiments
(17 types shown in Table~\ref{tab-applications}),
then construct the corresponding emulating bot scripts
for each user behavior as described in Section~\ref{sec-implementation},
and use them as a reference to label the time series data collected from victim smartphones.
We then shuffle these collected time series data with a one-time generated random number seed
and divide these data into a training and test set in a 7:3 ratio.
For 200 time-series data corresponding to each user behavior,
we randomly select and insert 140 time-series data into the training set
and the remaining 60 into the test set.
The data between the training and test sets will not be mixed in our subsequent experiments,
i.e., we will never use the data in the test set to train classification models.
Then, we use the training set to build the classification models and the test set to evaluate their accuracy.

\textbf{Notes.}
In our experiments, we only consider a total of 17 user behaviors for 4 applications.
So if the user is using another application or performing another behavior not included in our list,
the classifier will incorrectly classify it as one of them.
However, we do not consider this a significant issue since
the inference accuracy is still 0.8911 for 41 user behaviors shown in Section~\ref{sec-multi_behavior_trace}.
And if the behaviors considered are sufficient to cover the user's daily use,
the classifier will not be confused by other behaviors.

\subsection{Inference Accuracy of Fine-Grained User Behaviors\label{sec-accuracy}}
We first evaluate the accuracy of each application and the fine-grained user behaviors,
then analyze all four applications and their user behaviors together.
Afterward,
we evaluate the accuracy of two additional user behaviors,
i.e., the state of the background music and the PiP hover window.

\begin{figure}[!htbp]
\centering
\includegraphics[width=2.7in]{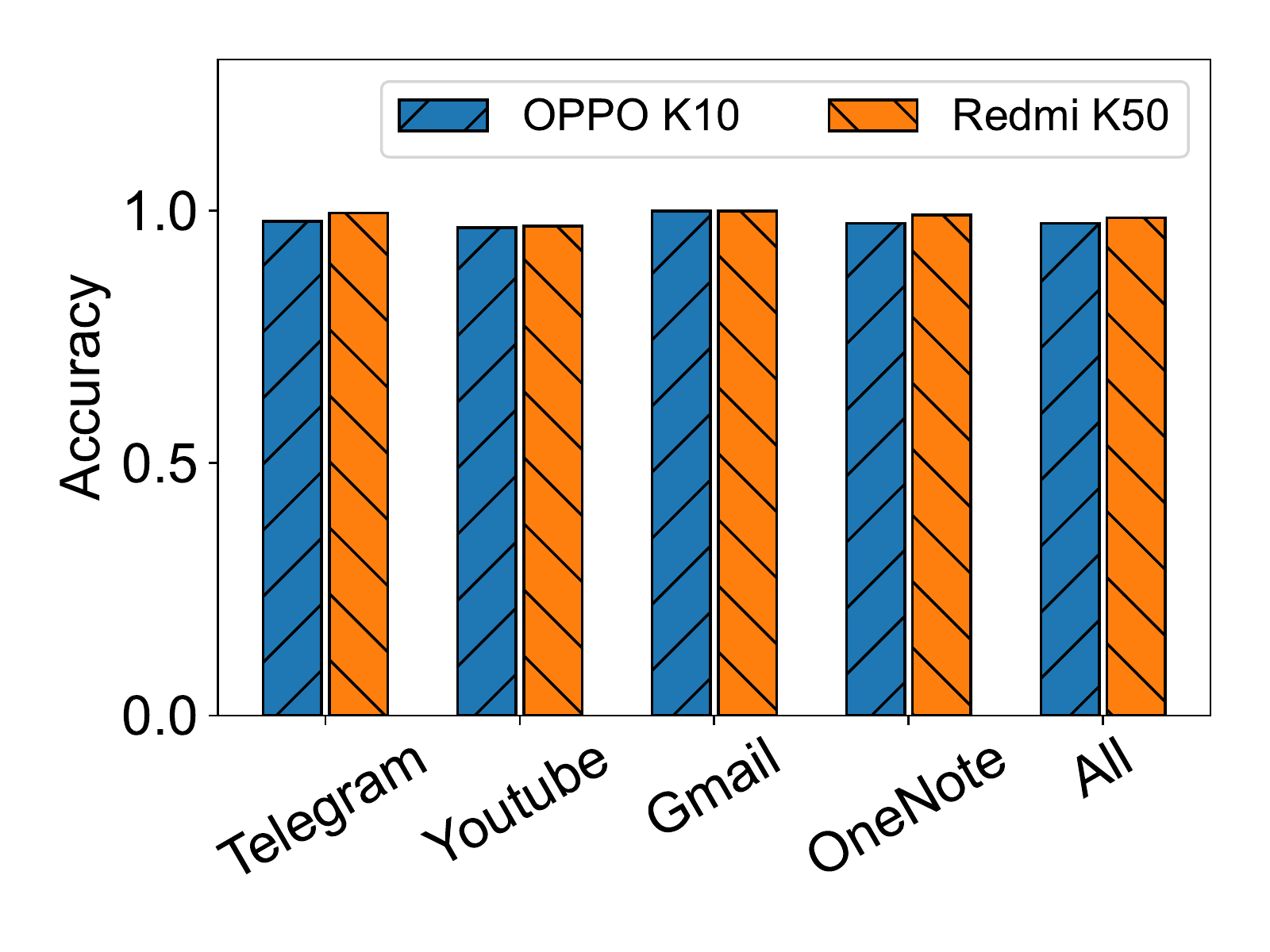}
\caption[]{Accuracy of inferring user behaviors on individual applications and all user behaviors.}
\label{fig-accuracy}
\end{figure}

\begin{figure}[!htbp]
\centering
\includegraphics[width=2.7in]{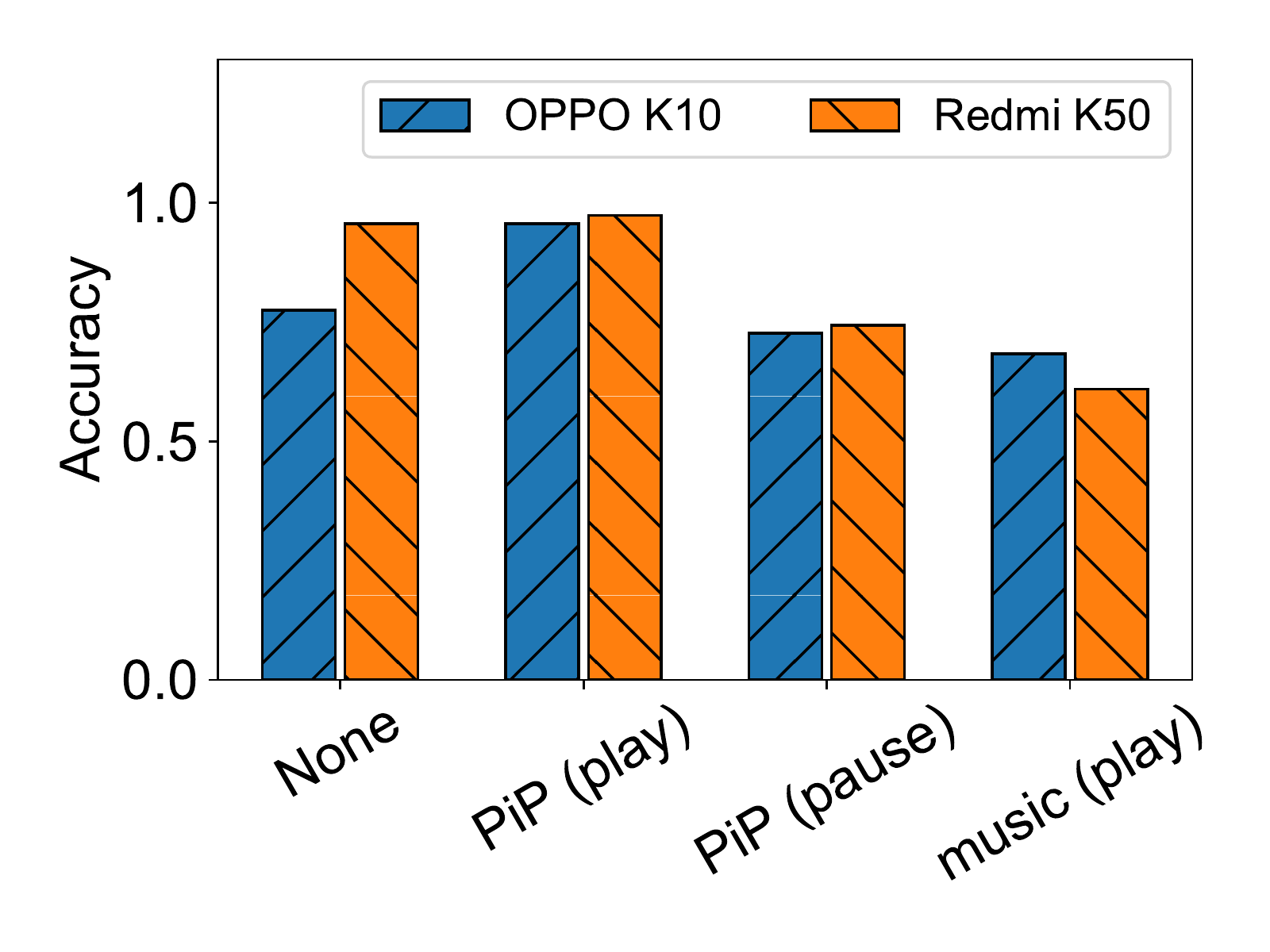}
\caption[]{Accuracy of inferring the state of the background music and the PiP hover window.}
\label{fig-accuracy-additional}
\end{figure}

\textbf{Accuracy of Individual Applications.}
First, we evaluate the accuracy of individual applications listed in Table~\ref{tab-applications}.
The result in Fig.~\ref{fig-accuracy} shows that our attack can successfully infer
user behaviors on individual applications with an accuracy above 96\%.
Meanwhile, we can identify the four user behaviors with 100\% accuracy for Gmail on both smartphones.

\textbf{Accuracy of All User Behaviors.}
Second, we combine all the user behaviors into one set, then evaluate the accuracy of this set.
Fig.~\ref{fig-accuracy} shows that the accuracy of identifying user behaviors from
the whole test set (17 different behaviors) can still reach 98\% on average.

\textbf{Accuracy of Additional Behaviors.}
Finally, we evaluate the accuracy of the state of the background music and the state of the PiP hover window.
Fig.~\ref{fig-accuracy-additional} shows that our attack can successfully infer these two behaviors
even when the user is performing other user behaviors on the foreground application,
and the accuracy of identifying these states is about 80\%.

\begin{figure*}[!htbp]
\begin{minipage}{2.35in}
\centering
\includegraphics[width=\linewidth]{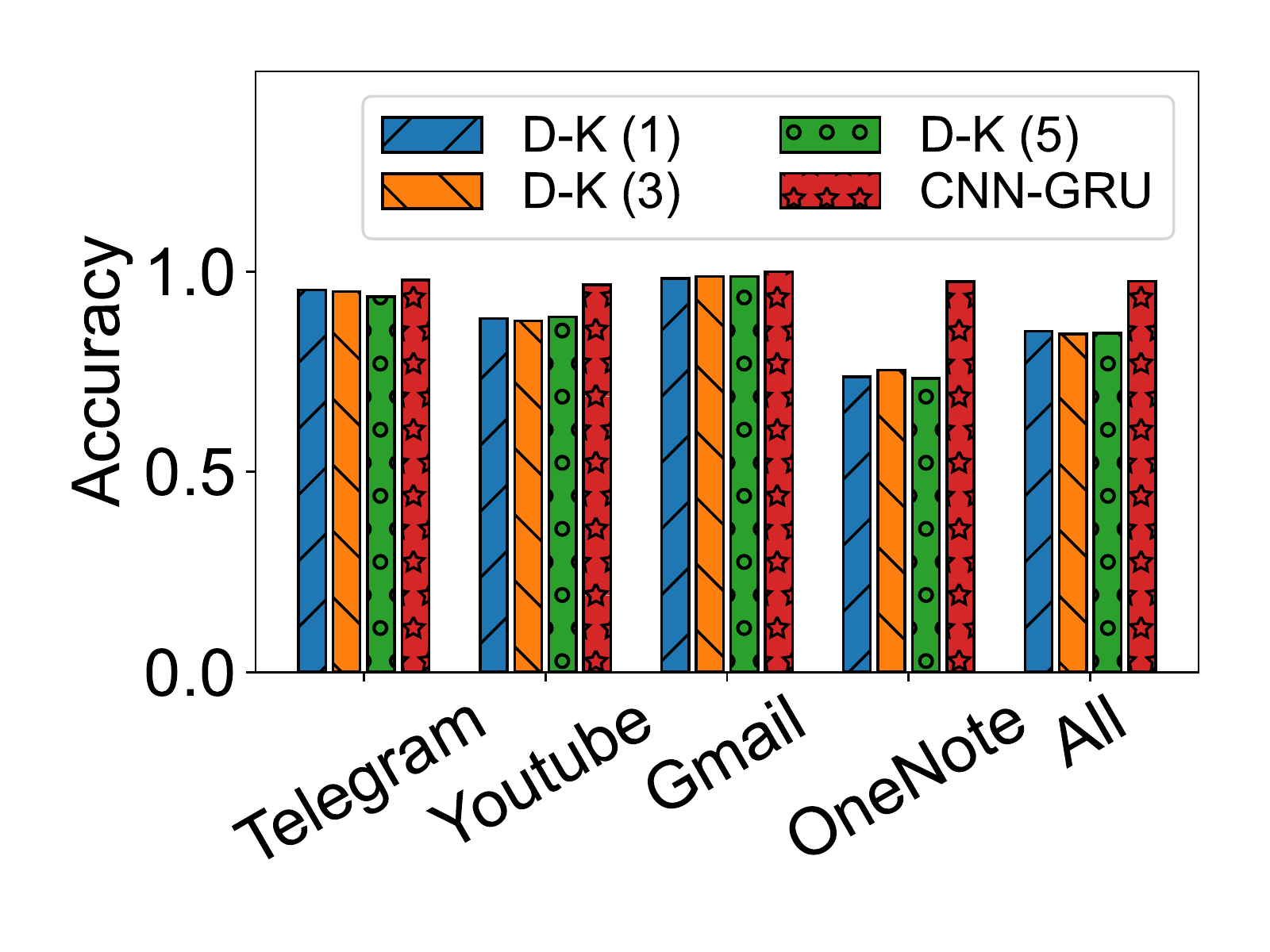}
\caption[]{Accuracy of using CNN-GRU network and DTW-KNN algorithm.}
\label{fig-impact-cnngru}
\end{minipage}
\hfill
\begin{minipage}{2.35in}
\centering
\includegraphics[width=\linewidth]{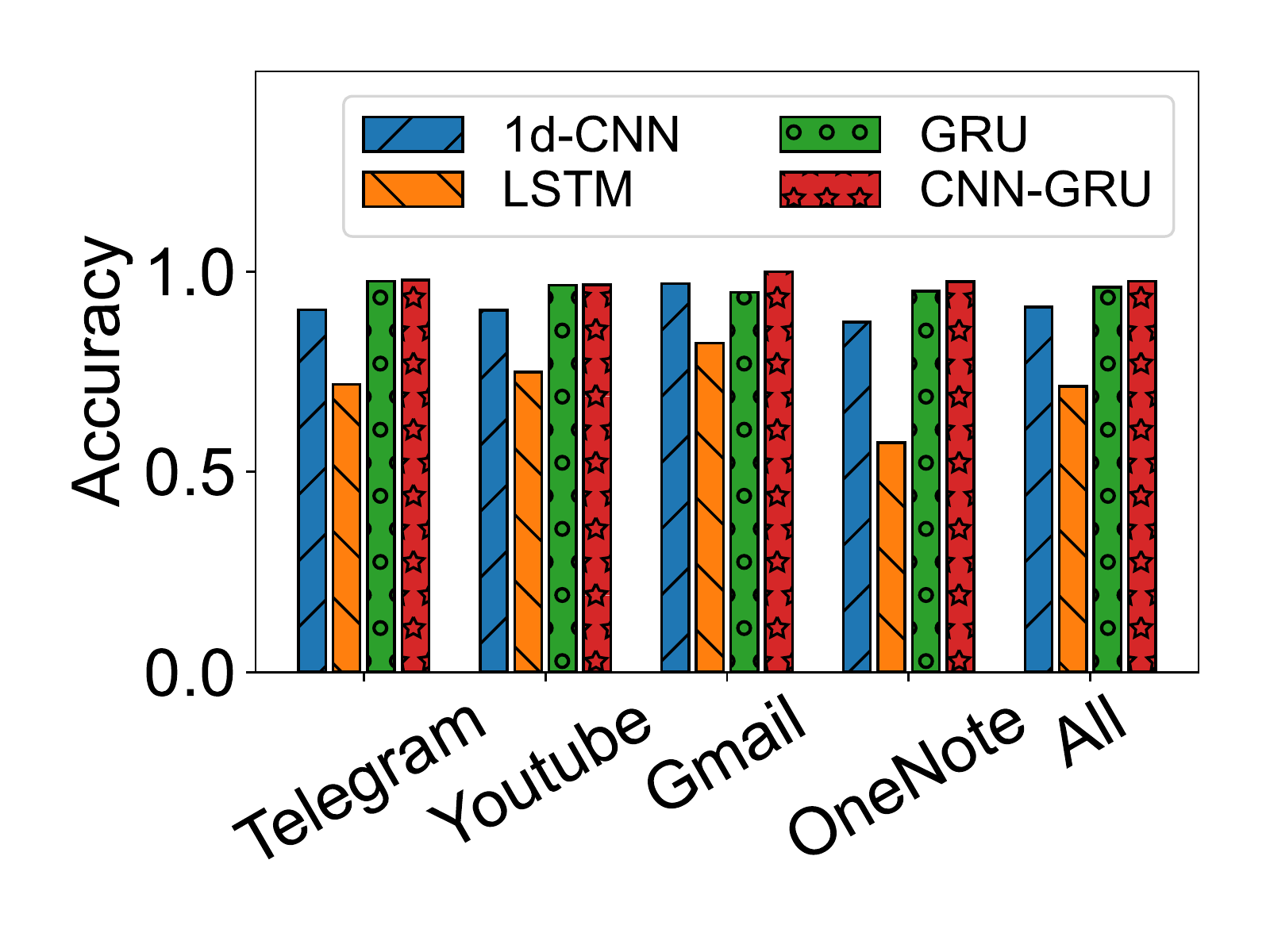}
\caption[]{Accuracy of using different deep learning methods.}
\label{fig-impact-cnngru2}
\end{minipage}
\hfill
\begin{minipage}{2.35in}
\centering
\includegraphics[width=\linewidth]{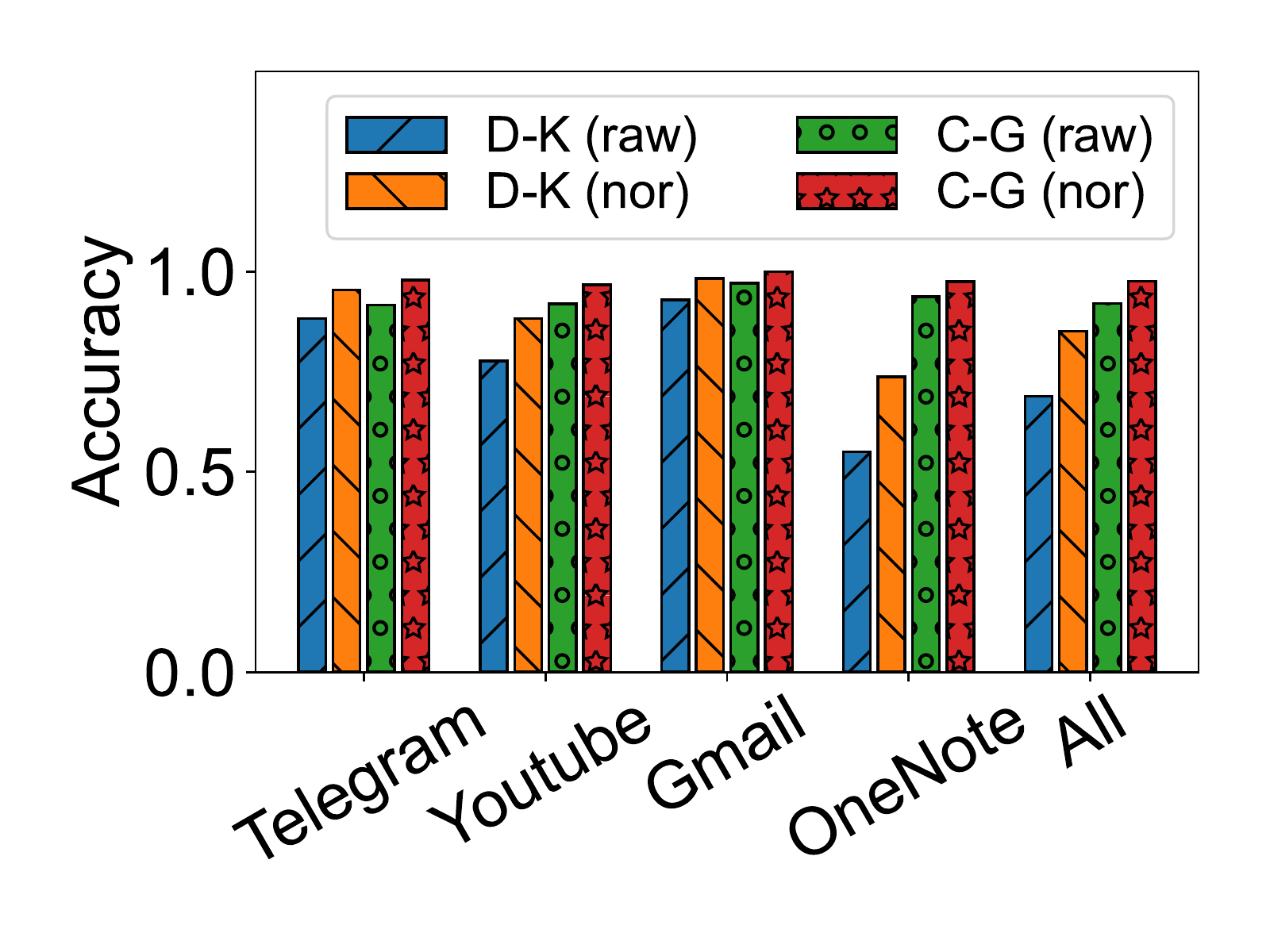}
\caption[]{Accuracy of using min-max normalized data and raw data.}
\label{fig-impact-minmax}
\end{minipage}
\\
\begin{minipage}{2.35in}
\centering
\includegraphics[width=\linewidth]{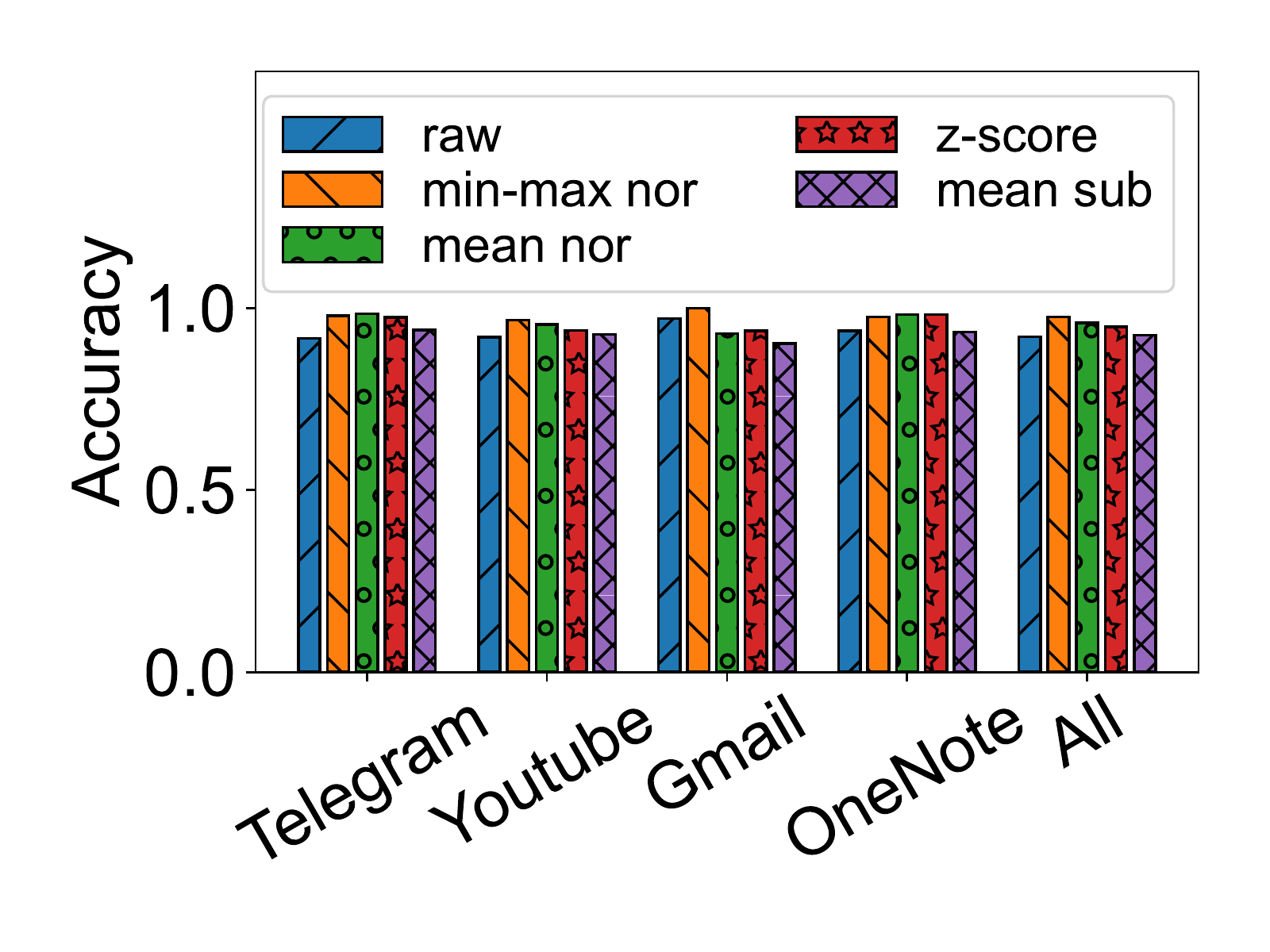}
\caption[]{Accuracy of using different data preprocessing methods.}
\label{fig-impact-minmax2}
\end{minipage}
\hfill
\begin{minipage}{2.35in}
\centering
\includegraphics[width=\linewidth]{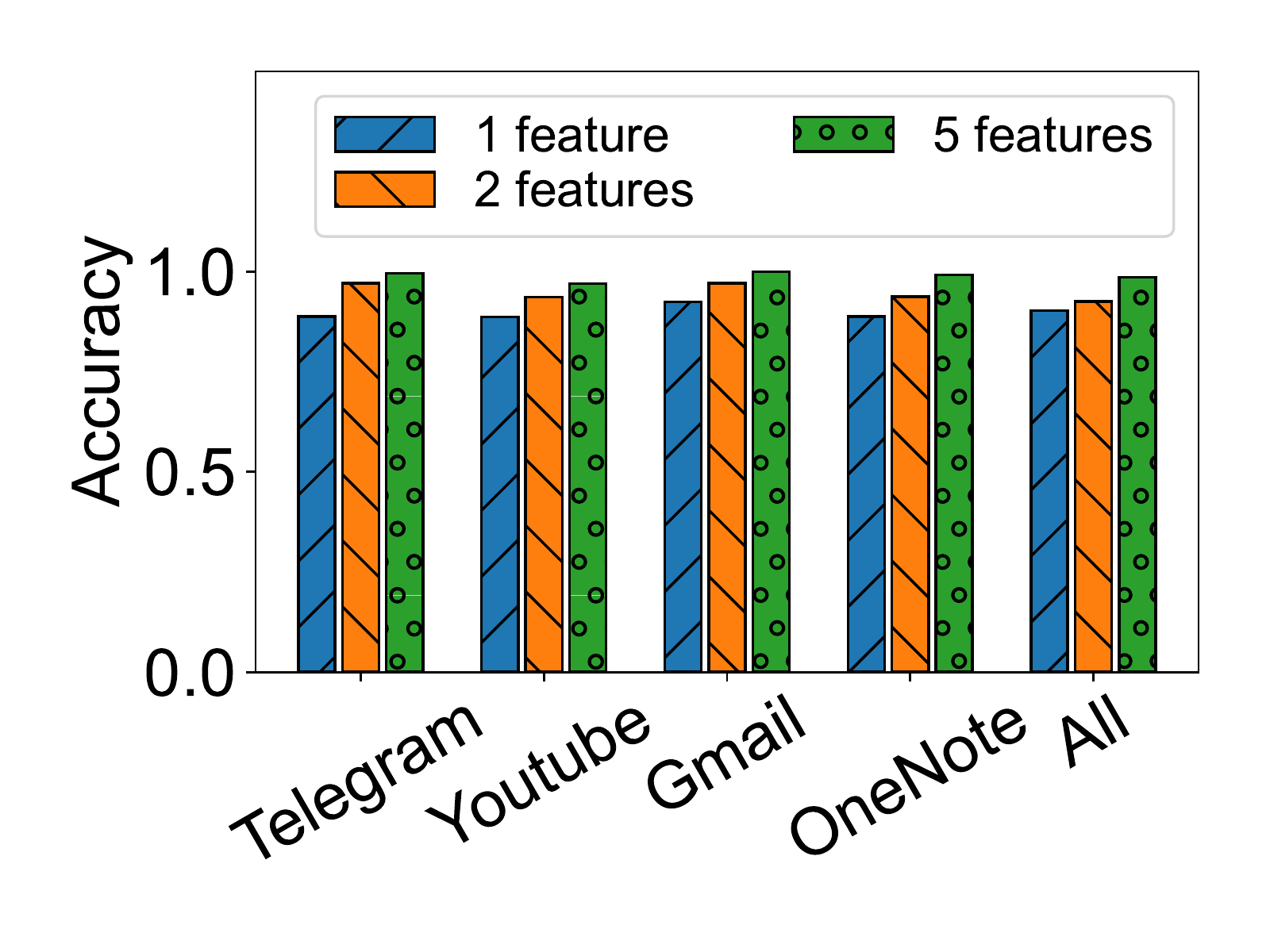}
\caption[]{Accuracy of using different feature dimensions.}
\label{fig-impact-dimension}
\end{minipage}
\hfill
\begin{minipage}{2.35in}
\centering
\includegraphics[width=\linewidth]{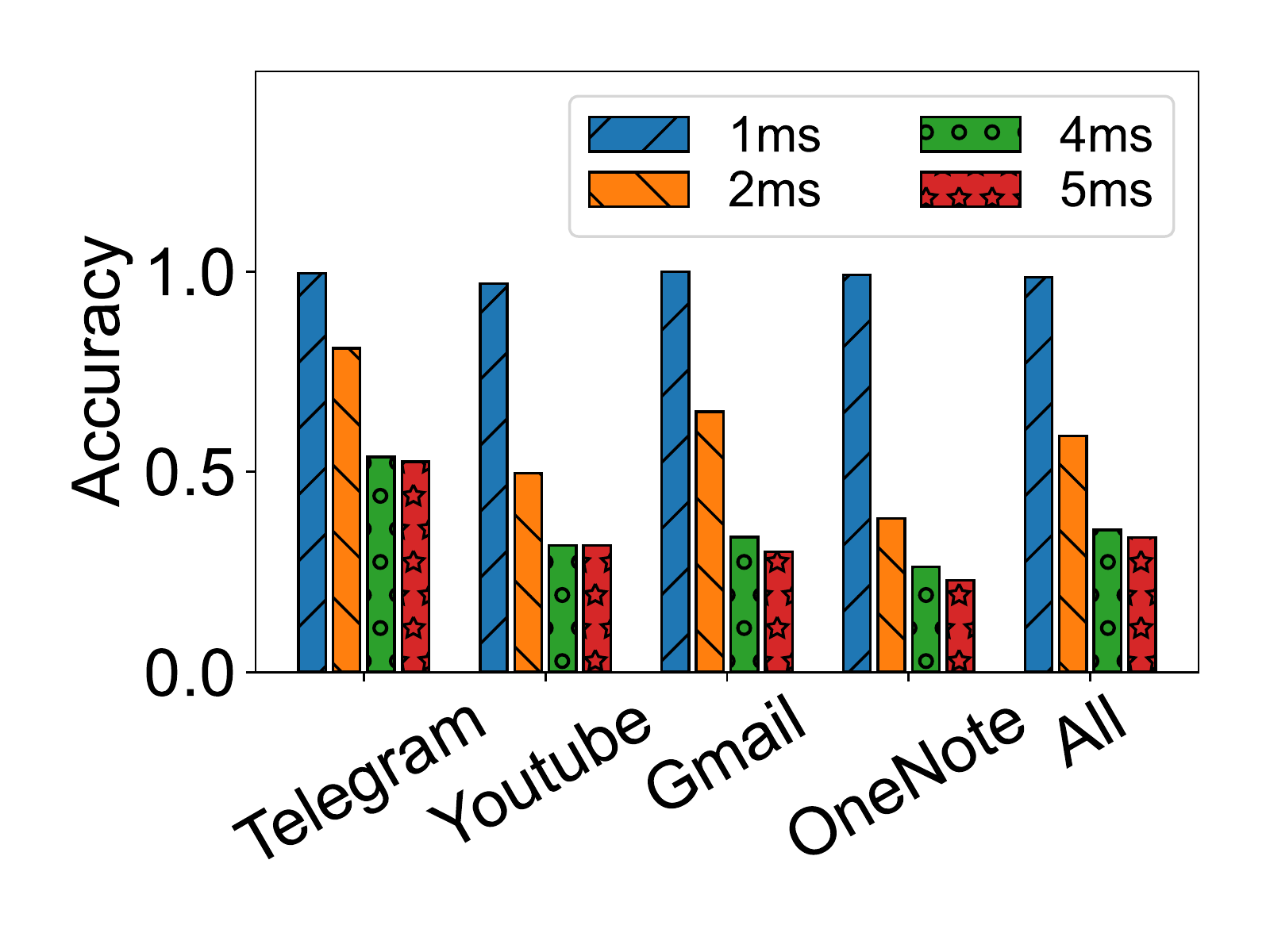}
\caption[]{Accuracy of using different invoking intervals.}
\label{fig-impact-interval}
\end{minipage}
\end{figure*}

\subsection{Evaluation of Impact Factors}
\subsubsection{Impact of CNN-GRU Network\label{sec-impact-cnn-gru}}
To evaluate the impact of our CNN-GRU network on the accuracy of our attack,
we first use the DTW-KNN algorithm (with different values of $K$) and the CNN-GRU network
to separately classify user behaviors on OPPO K10.
We also process the data using the min-max normalization algorithm.
As shown in Fig.~\ref{fig-impact-cnngru},
the accuracy of the CNN-GRU network is 15\% higher than the accuracy of
the DTW-KNN algorithm with the best $K$ value
(0.9755 for the CNN-GRU network and 0.8510 for the DTW-KNN algorithm).
In particular, the gap in accuracy between the two is up to 29\%
(0.9750 and 0.7542) for OneNote.

Then, we also compare the accuracy of the CNN-GRU network with alone CNN and RNN models.
The CNN model used in our experiment is the 1D-CNN network~\cite{tang2020rethinking},
and the RNN model used in our experiment is an LSTM-only network and a GRU-only network.
Fig.~\ref{fig-impact-cnngru2} shows that the accuracy of the CNN-GRU model is
7\% higher than the accuracy of the 1D-CNN model (0.9755 and 0.9118),
37\% higher than the accuracy of the LSTM model (0.9755 and 0.7137),
and 1.5\% higher than the accuracy of the GRU model (0.9755 and 0.9608).
The result demonstrates that both CNN and RNN models can be used to classify user behaviors,
but our CNN-GRU model can achieve the best accuracy.

\subsubsection{Impact of Min-Max Normalization\label{sec-impact-minmax}}
We evaluate the impact of the min-max normalization algorithm on the accuracy of our attack through:
\begin{itemize}
\item Comparing the inference accuracy between normalized data and raw data;
\item Comparing the inference accuracy among four different preprocessing algorithms.
\end{itemize}
And both experiments demonstrate that the min-max normalization algorithm can improve the accuracy of our attack.

First, we use the raw time series data and the min-max normalized data separately to
build the classification models using the DTW-KNN algorithm and our CNN-GRU network,
then evaluate their accuracy on OPPO K10.
The result in Fig.~\ref{fig-impact-minmax} shows that
the accuracy of using min-max normalized data is 23\% higher than the accuracy
of using raw data for the DTW-KNN algorithm (0.6892 for raw data and 0.8510 for min-max normalized data)
and 6\% higher for the CNN-GRU network (0.9206 for raw data and 0.9755 for min-max normalized data).

Then, we infer the user behaviors using four different data preprocessing methods
(min-max normalization, mean normalization, z-score standardization, and mean subtraction)
on OPPO K10 with our CNN-GRU network.
As shown in Fig.~\ref{fig-impact-minmax2},
all the preprocessing methods can improve the accuracy of our attack compared with the baseline (0.9206).
And the accuracy of using min-max normalization ranks first among all the preprocessing approaches (0.9755),
which is 1.7\% higher than the accuracy of using mean normalization (0.9588),
2.9\% higher than the accuracy of using z-score standardization (0.9480),
and 5.4\% higher than the accuracy of using mean subtraction (0.9255) on average.

\subsubsection{Impact of Feature Dimension\label{sec-impact-feature}}
To evaluate the impact of feature dimension on the accuracy of our attack,
we use our CNN-GRU network with different feature dimensions (1, 2, and 5) to classify user behaviors on Redmi K50.
And we also use the data preprocessed using the min-max normalization algorithm to evaluate the accuracy.
Since there are differences in the consumption of resources such as storage and memory by different user behaviors,
and different return values have different preferences for identifying different user behaviors,
we select the feature or combination of features with the highest accuracy for feature fusion.

From the result of our experiments shown in Fig.~\ref{fig-impact-dimension},
we can see that the more features we use, the higher the accuracy of our attack.
When we use only one feature, the accuracy of our attack is 0.9029,
and when we use 5 features, the accuracy of our attack reaches 0.9863,
which is 9\% higher than the accuracy of using only one feature.
Although the model with feature fusion leads to worse results than the model without feature fusion in some cases,
the experimental results show that our chosen return values all have a positive effect.

\subsubsection{Impact of Invoking Interval\label{sec-impact-interval}}
The speed of invoking system call return values affects the accuracy of identifying user behaviors.
To evaluate the impact of the interval of invoking the system call return values,
we analyze the accuracy of our attack with different invoking intervals (1ms, 2ms, 4ms, and 5ms) on Redmi K50,
and the inference accuracy in these cases is shown in Fig.~\ref{fig-impact-interval}.
The accuracy of inferring user behaviors decreases by 40\% (from 0.9863 to 0.5902)
when the invoking interval increases from 1ms to 2ms,
and by 66\% (from 0.9863 to 0.3363) when the invoking period increases to 5ms.
Therefore, we recommend setting the invoking interval to no more than 2ms for accurate inference.

\begin{figure}[!htbp]
\begin{minipage}{0.49\linewidth}
\centering
\includegraphics[width=\linewidth]{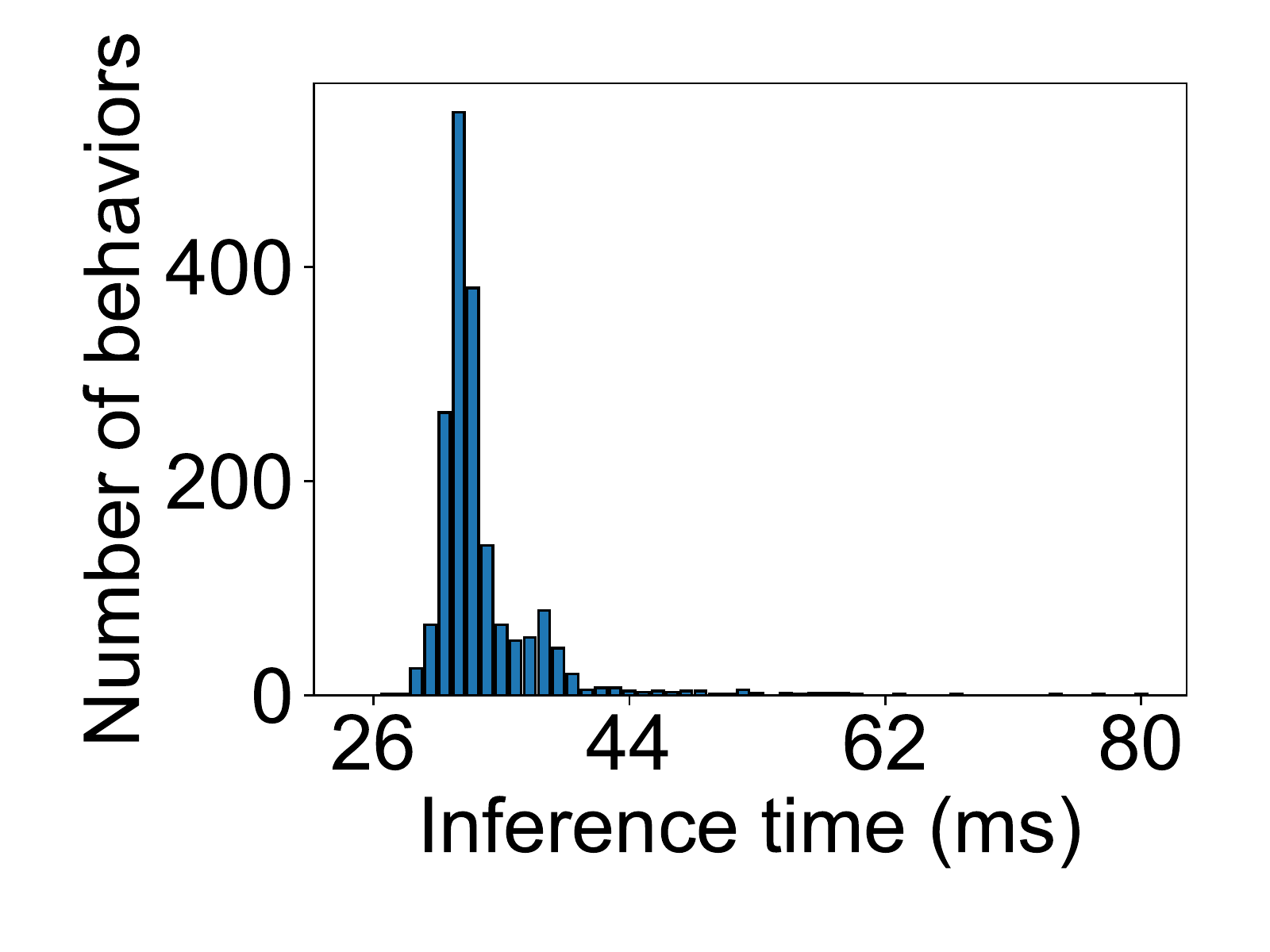}
\caption[]{Timeliness evaluation of our attack.}
\label{fig-overhead-time}
\end{minipage}
\hfill
\begin{minipage}{0.49\linewidth}
\centering
\includegraphics[width=\linewidth]{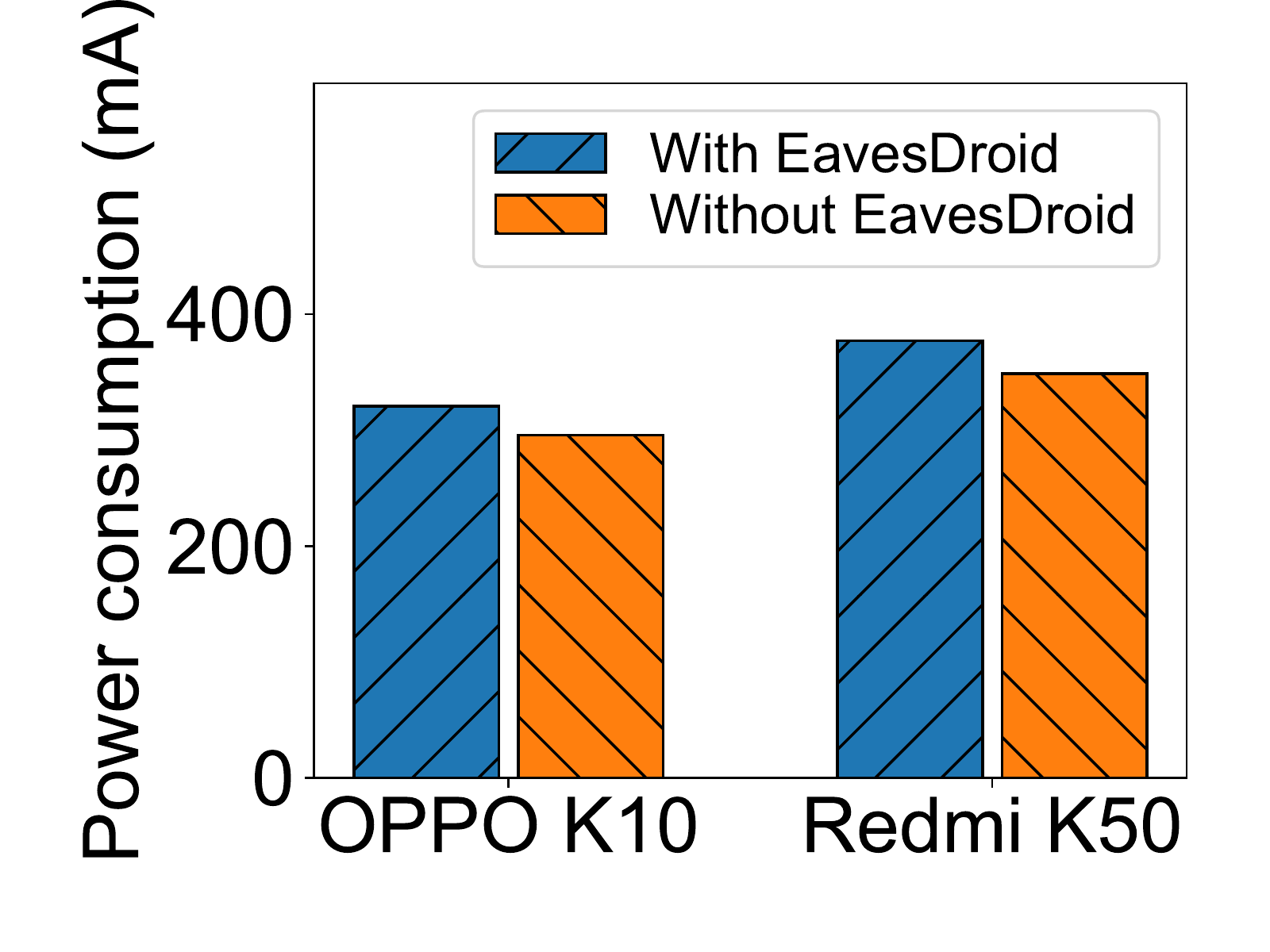}
\caption[]{Power consumption evaluation of our attack.}
\label{fig-overhead-power}
\end{minipage}
\\
\begin{minipage}{0.49\linewidth}
\centering
\includegraphics[width=\linewidth]{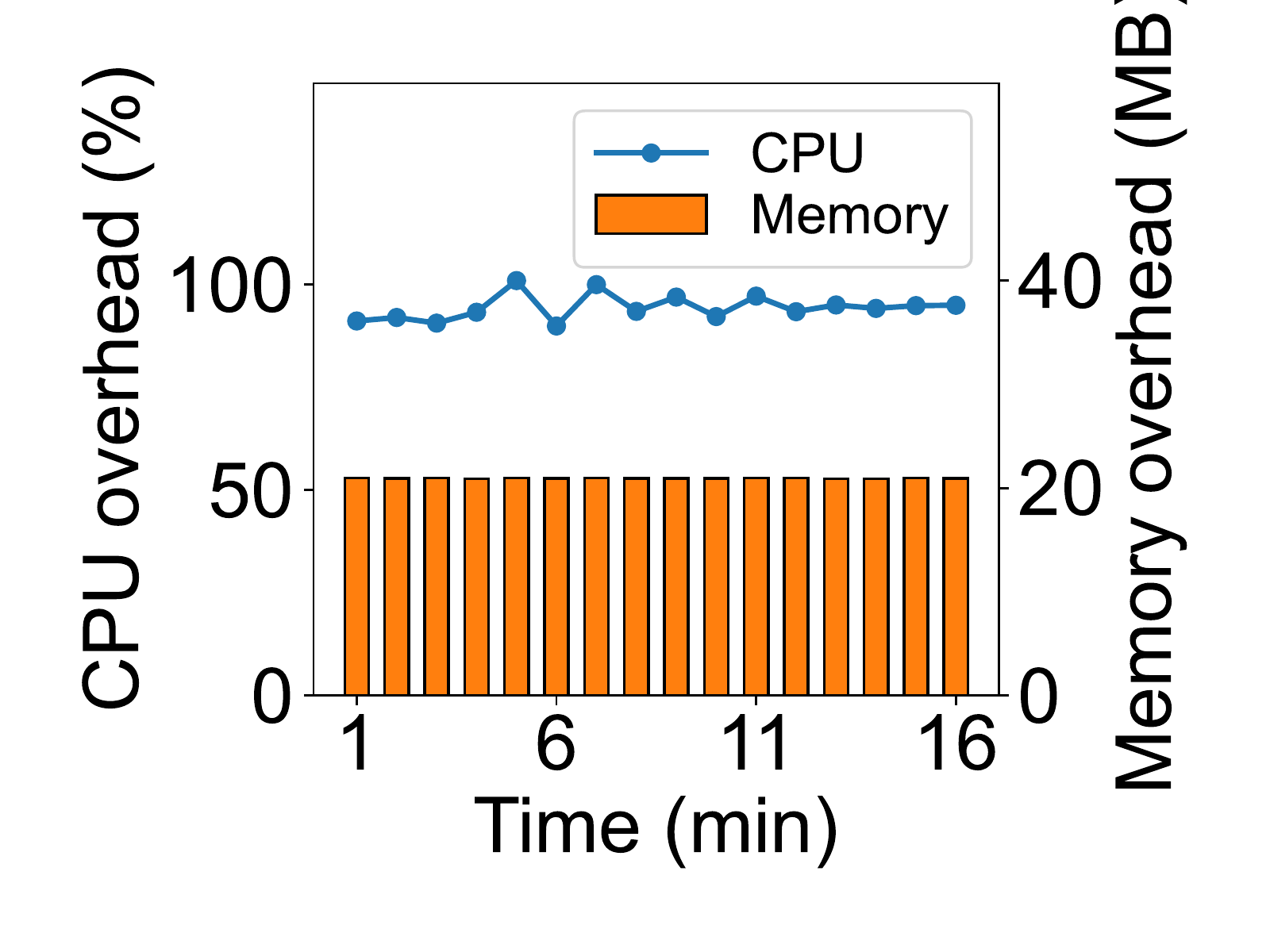}
\caption[]{Computational resource consumption evaluation of our attack.}
\label{fig-overhead-compute}
\end{minipage}
\hfill
\begin{minipage}{0.49\linewidth}
\centering
\includegraphics[width=\linewidth]{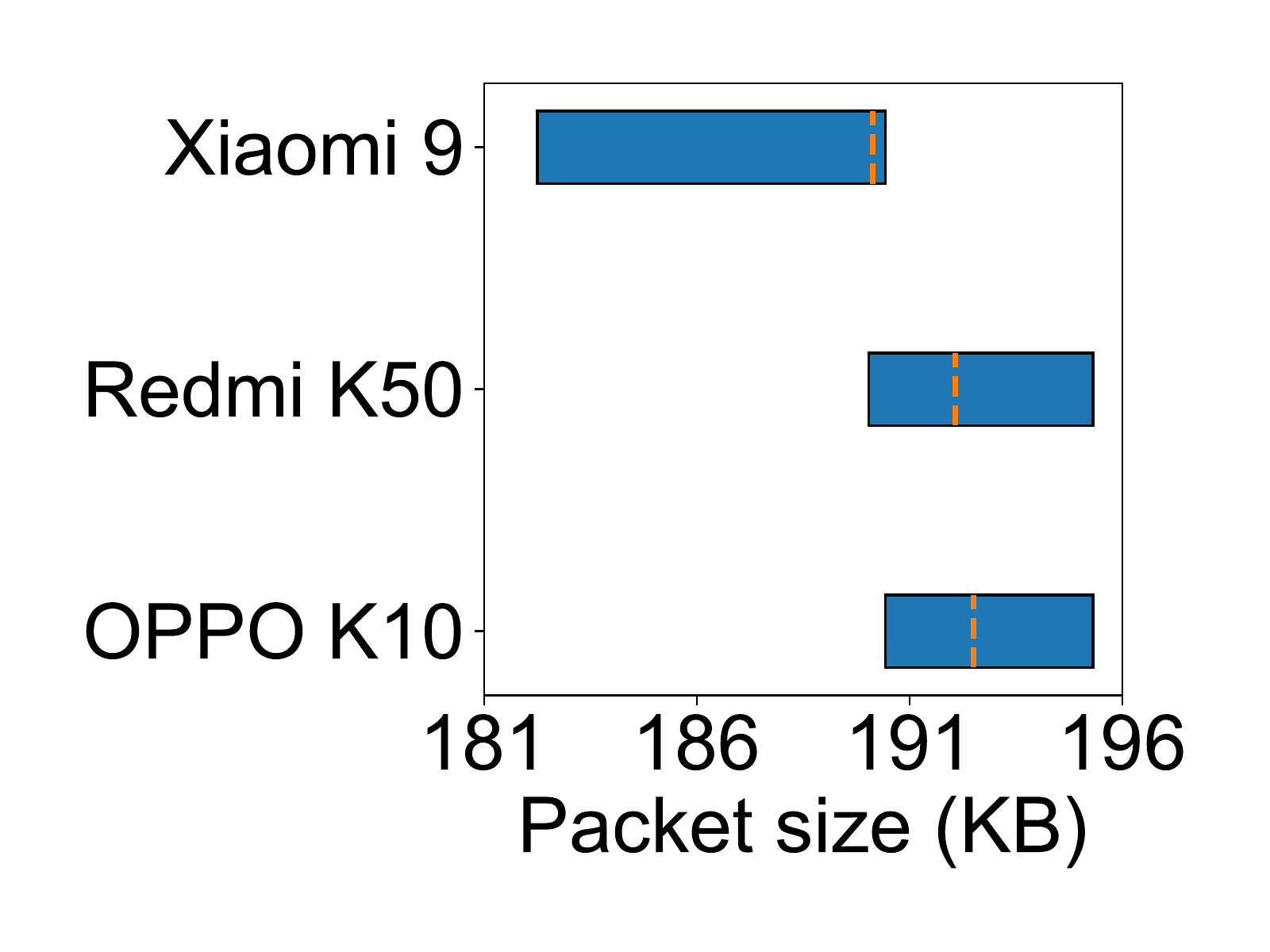}
\caption[]{Network traffic evaluation of our attack.}
\label{fig-overhead-traffic}
\end{minipage}
\end{figure}

\subsection{Timeliness and Overhead}
\textbf{Timeliness.}
First, we evaluate the timeliness of our attack.
In our analysis, we ignore the time delay caused by network transmission and
use the time from when the attacker receives the uploaded data from the victim's smartphone
to the output of the CNN-GRU classifier as the evaluation metric for timeliness.
Fig.~\ref{fig-overhead-time} shows the time needed to identify 1800 fine-grained user behaviors on OPPO K10.
The result demonstrates that the average time is 33.63ms, and over 95\% of behaviors can be identified in 40ms.

\textbf{Power Consumption.}
Next, we evaluate the power consumption of our attack on smartphones.
We use AccuBattery to collect the electric current data when the victim's smartphone is idle
and when the victim's smartphone is running our attack application.
The result shown in Fig.~\ref{fig-overhead-power} demonstrates that our attack application consumes 8\% additional power on average.
When the application is not used, the average power consumption is about 322mA,
while when it is used, the average power consumption is about 349mA.

\textbf{Computational Resource Consumption.}
Afterward, we evaluate the computational resource consumption of our attack on the OnePlus 7Pro.
In this experiment, we use a monitoring application called Scene,
which can read the CPU and memory usage of any running application
to evaluate computational resource consumption.
We run our attack application on the smartphone for 15 minutes and monitor the CPU and memory usage.
As shown in Fig.~\ref{fig-overhead-compute}, the average CPU usage is 94.4\% on a single core,
and the average memory usage is 20.94 MB when using our attack application.

\textbf{Traffic Overhead.}
Finally, we evaluate the network traffic overhead of our attack by analyzing data uploaded from several smartphones.
As shown in Fig.~\ref{fig-overhead-traffic}, the average packet size is 191.6 KB.
The blue bars represent the packet size range on each device, and the orange dashed lines represent the average packet size.

\subsection{Adaptability of Attack}
We evaluate the adaptability of our attack through
cross-device experiments and version-drift experiments.

\subsubsection{Cross-Device Experiments}
\begin{figure}[!htbp]
\centering
\includegraphics[width=2.7in]{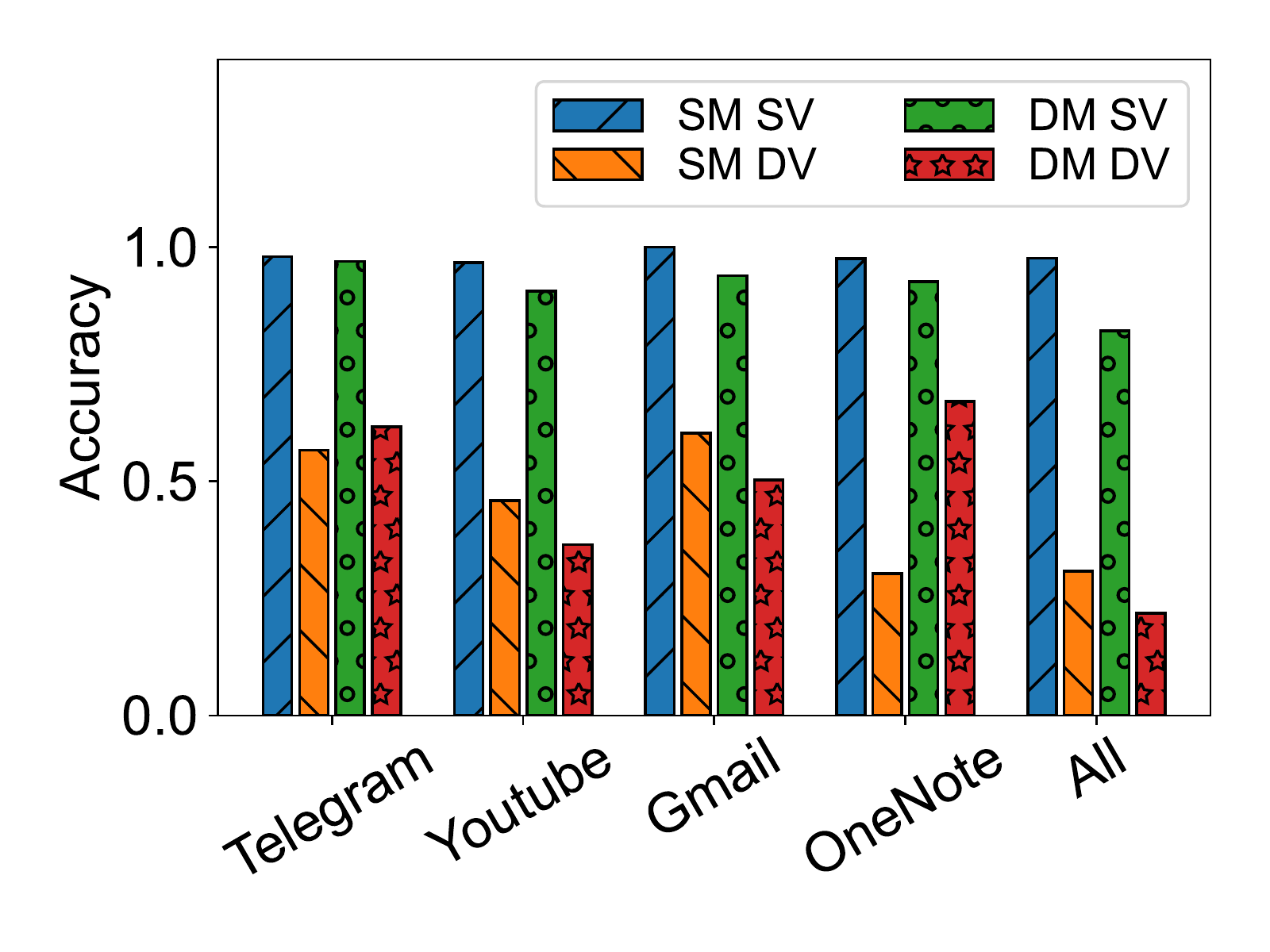}
\caption[]{Cross-device adaptability evaluation of our attack.
SM SV: same model and same version;
SM DV: same model and different versions;
DM SV: different models and same version;
DM DV: different models and different versions.}
\label{fig-adaptability}
\end{figure}

We perform four experiments to evaluate
the inference accuracy across different Android device models and versions,
and the devices and versions are listed in Table~\ref{tab-adaptability}.
The result shown in Fig.~\ref{fig-adaptability} demonstrates that
the accuracy of our attack is up to 0.9755 for devices with the same model and the same Android version,
0.8212 for devices with different models and the same Android version,
both of which indicate that our attack is adaptable to devices with the same Android version.
However, the accuracy of our attack drops to 0.3076 for devices with the same model and different Android versions,
0.2183 for devices with different models and Android versions.
Therefore, We cannot successfully launch our attack on devices with different Android versions,
and we also find that the same user behavior can cause various return value changes for different Android versions,
which may be due to the diverse execution mechanisms of user programs under different Android versions.

\begin{table}[!htbp]
\centering
\caption[]{Device Information Used to Evaluate the Cross-Device Adaptability of Our Attack and Evaluation Results}
\label{tab-adaptability}
\begin{tabular}{cccc}
\toprule
Device A & Device B & Description & Accuracy \\
\midrule
\makecell[c]{OPPO K10 \\ (Android 12)} & \makecell[c]{OPPO K10 \\ (Android 12)}
& \makecell[c]{same model \\ same version} & 0.9755 \\
\midrule
\makecell[c]{Xiaomi 9 \\ (Android 11)} & \makecell[c]{Xiaomi 9 \\ (Android 12)}
& \makecell[c]{same model \\ different versions} & 0.3076 \\
\midrule
\makecell[c]{OPPO K10 \\ (Android 12)} & \makecell[c]{Redmi K50 \\ (Android 12)}
& \makecell[c]{different models \\ same version} & 0.8212 \\
\midrule
\makecell[c]{Redmi K50 \\ (Android 12)} & \makecell[c]{Xiaomi 9 \\ (Android 11)}
& \makecell[c]{different models \\ different versions} & 0.2183 \\
\bottomrule
\end{tabular}
\end{table}

\subsubsection{Version-Drift Experiments}
We also evaluate the inference accuracy of our attack on the same application with different versions.
The experiment is conducted on Xiaomi 9,
and the application versions are listed in Table~\ref{tab-adaptability2}.
The experimental result shows that our attack can handle the version drift of applications.
The inference accuracy of our attack is up to 0.9925 and 0.9825 for each application with the same version,
still about 0.8475 and 0.8537 for each application with high version drift.
We believe that our attack can handle version drift because
the operating logic and UI of applications are relatively stable across different versions.

\begin{table}[!htbp]
\centering
\caption[]{Application and Version Information Used to Evaluate the Version Drift of Our Attack, and Evaluation Results}
\label{tab-adaptability2}
\begin{tabular}{cccc}
\toprule
Application & Model Version & Test Version & Accuracy \\
\midrule
\multirow{6}{*}{Telegram} & \multirow{6}{*}{9.0.0}
&   8.7.0 & 0.8475 \\ %
\cmidrule{3-4}
& & 8.8.2 & 0.8763 \\ %
\cmidrule{3-4}
& & 8.9.0 & 0.9762 \\ %
\cmidrule{3-4}
& & 9.0.0 & 0.9925 \\ %
\midrule
\multirow{9}{*}{Gmail} & \multirow{9}{*}{\makecell[c]{2022.08.07 \\ (468496502)}}
&   \makecell[c]{2021.09.19 \\ (399766500)} & 0.8537 \\ %
\cmidrule{3-4}
& & \makecell[c]{2022.01.09 \\ (422969509)} & 0.9050 \\ %
\cmidrule{3-4}
& & \makecell[c]{2022.05.15 \\ (454752110)} & 0.9287 \\ %
\cmidrule{3-4}
& & \makecell[c]{2022.08.07 \\ (468496502)} & 0.9825 \\ %
\bottomrule
\end{tabular}
\end{table}

\subsection{Evaluation in Complex Scenarios}
While our previously emulated behaviors are performed in a less-noisy and controlled environment,
these behaviors can be affected by other factors in the real world,
such as other running processes and changes in user behavior patterns
(e.g., the varying message length, the varying interval between keyboard input and the press of the send button).
Therefore, we also evaluate our attack in more complex scenarios.

\subsubsection{Multi-Behavior Traces\label{sec-multi_behavior_trace}}
First, when the user uses the smartphone,
he/she usually performs a series of behaviors in succession,
rather than only a specific behavior during our collection time (5s).
Therefore, we evaluate the inference accuracy of our attack on multi-behavior traces on Xiaomi 9.

In our experiment, we set the number of user behaviors in each trace from 1 to 3.
There are 17 categories of one-behavior traces,
12 categories of two-behavior traces,
and 12 categories of three-behavior traces.
The experimental result shows the inference accuracy
is about 0.8911 for 41 categories of behavior traces.
Among them, the accuracy is 0.8706, 0.9287, and 0.8833 for single behaviors,
two-behavior combinations, and three-behavior combinations,
which indicates that our attack can be successfully launched with multi-behavior traces.

\subsubsection{Noisy Execution Environment}
Second, in our previous experiments, user behaviors are emulated in a comparatively simple environment
with only system noise and no additional noise from other processes.
However, in real-world situations, many other processes are running in the background
that may affect the inference accuracy of our attack.
Therefore, we evaluate the robustness of our attack against noisy execution environments
by adding additional workloads along with the emulated behaviors on Xiaomi 9.

\begin{figure*}[!htbp]
\begin{minipage}{2.35in}
\centering
\includegraphics[width=\linewidth]{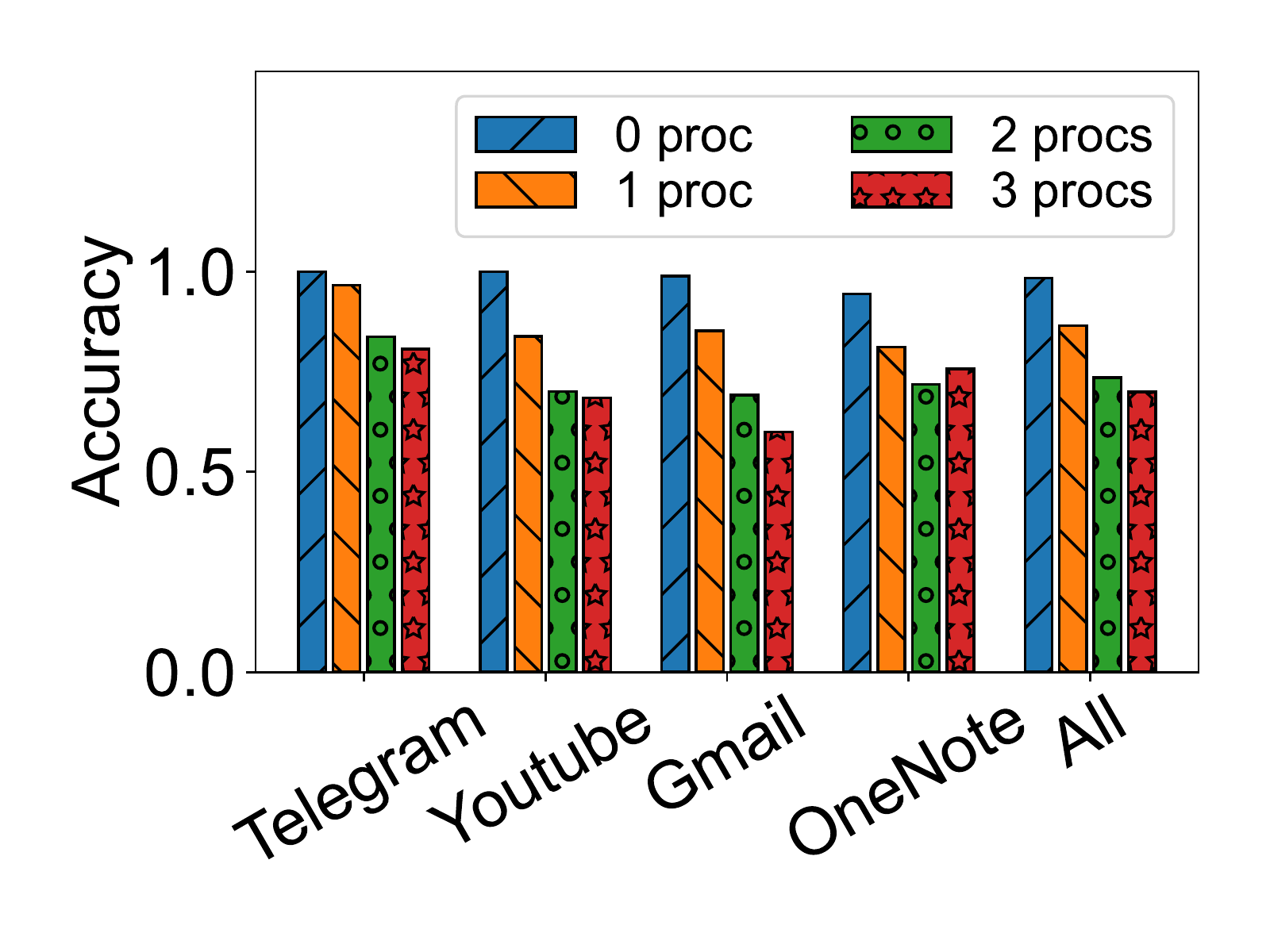}
\caption[]{Evaluation result of our attack in noisy execution scenarios (model without noisy data).}
\label{fig-noise}
\end{minipage}
\hfill
\begin{minipage}{2.35in}
\centering
\includegraphics[width=\linewidth]{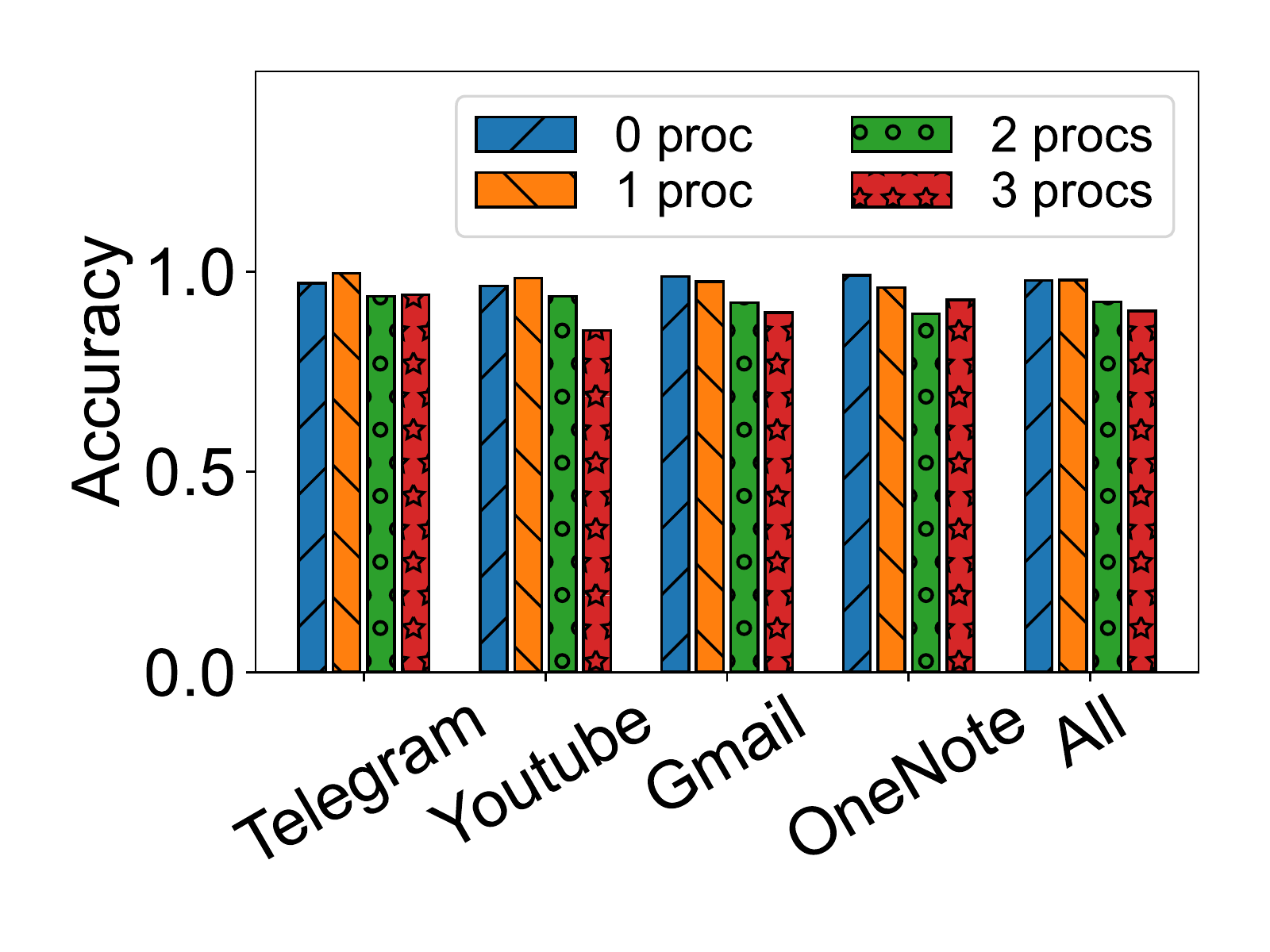}
\caption[]{Evaluation result of our attack in noisy execution scenarios (model with noisy data).}
\label{fig-noise2}
\end{minipage}
\hfill
\begin{minipage}{2.35in}
\centering
\includegraphics[width=\linewidth]{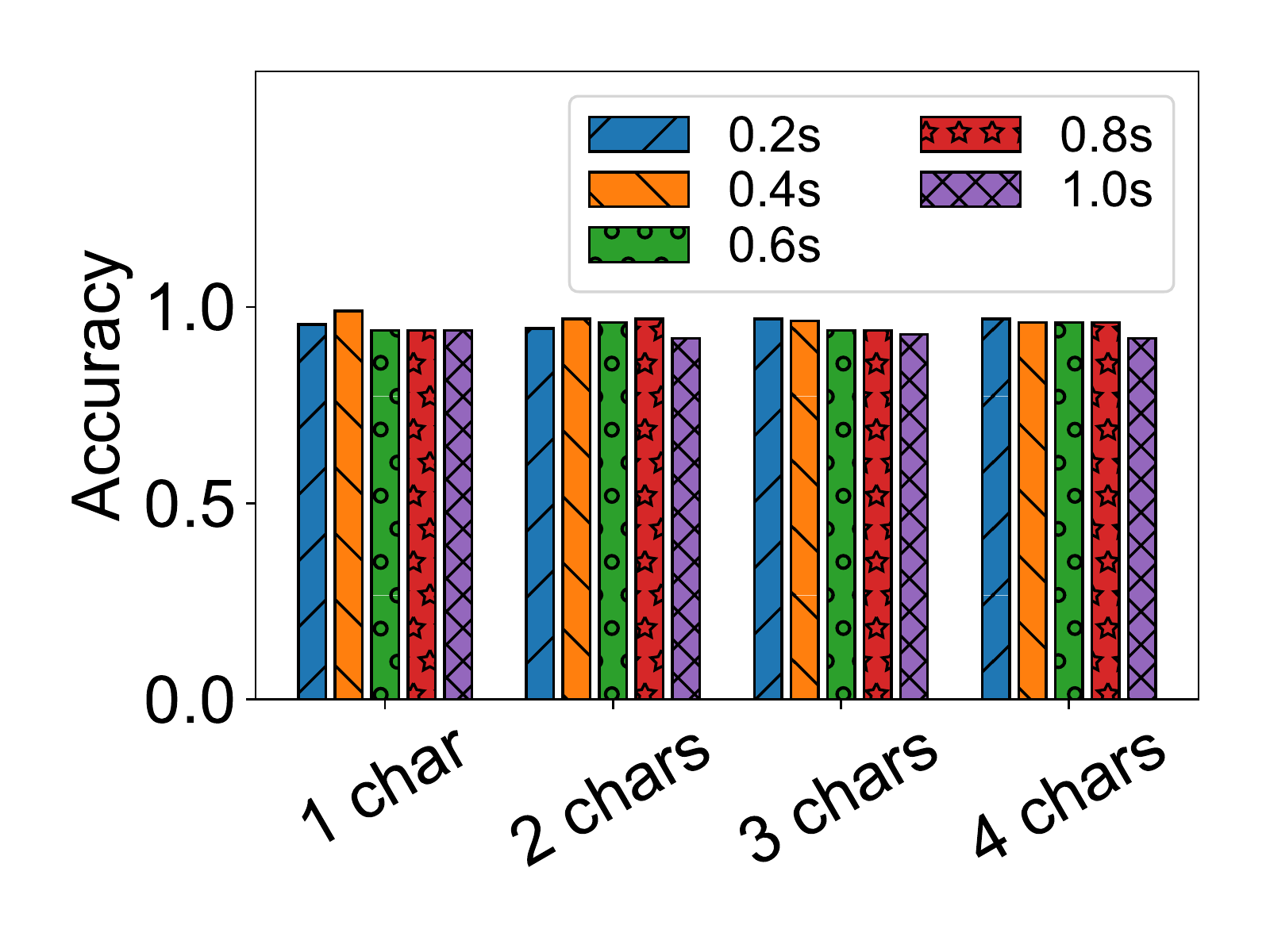}
\caption[]{Evaluation result of our attack against varied message lengths and operation intervals.}
\label{fig-pattern}
\end{minipage}
\end{figure*}

In our experiment, the number of additional running processes
that interfere with our attack ranges from 0 to 3.
The result in Fig.~\ref{fig-noise} demonstrates that our attack is
robust against noisy execution environments for the classification model
trained with no background noise.
The accuracy of our attack is as high as 0.9841 with 0 other running processes
and still 0.8653 with one other running process.
However, the inference accuracy drops to 0.7350 and 0.6993 when the number of
other running processes increases to 2 and 3, respectively,
which may challenge the robustness of our attack.

Therefore, to further improve and show the robustness of our attack,
we train another classification model with these noisy data and evaluate the inference accuracy
for the whole dataset with 0, 1, 2, and 3 additional running processes.
Fig.~\ref{fig-noise2} shows the attack results of our classification model trained on the noisy data,
with an inference accuracy of over 97\% for 0 and 1 other running processes.
While with 2 and 3 other running processes,
the inference accuracy drops slightly to 0.9245 and 0.9020,
indicating that our attack is still robust to the noisy execution environment
when we train the classification model with enough noisy data.

\subsubsection{Varied Message Lengths and Operation Intervals}
Third, due to varied message lengths, varied intervals between each keyboard input, and other reasons,
user behavior patterns sometimes differ in practical situations.
Therefore, we evaluate the inference accuracy of our attack by
varying the message length and the operation interval of sending messages on Xiaomi 9.
We vary the message length from 1 to 4 characters and the interval between keyboard input,
keyboard input and send button press from 0.2s to 1s (0.2s step).
As shown in Fig.~\ref{fig-pattern},
our attack can still infer the correct user behavior (i.e., input and send messages)
from 17 total behaviors with an inference accuracy above 0.9200,
indicating that our attack can be successfully launched
with varied message lengths and operation intervals.

\subsection{Stealthiness of Attack\label{sec-stealthiness}}
To evaluate the stealthiness of our attack,
we analyze our malicious application through static and run-time malware detection tools.

\subsubsection{Static Detection}
\begin{figure}[!htbp]
\centering
\includegraphics[width=\linewidth]{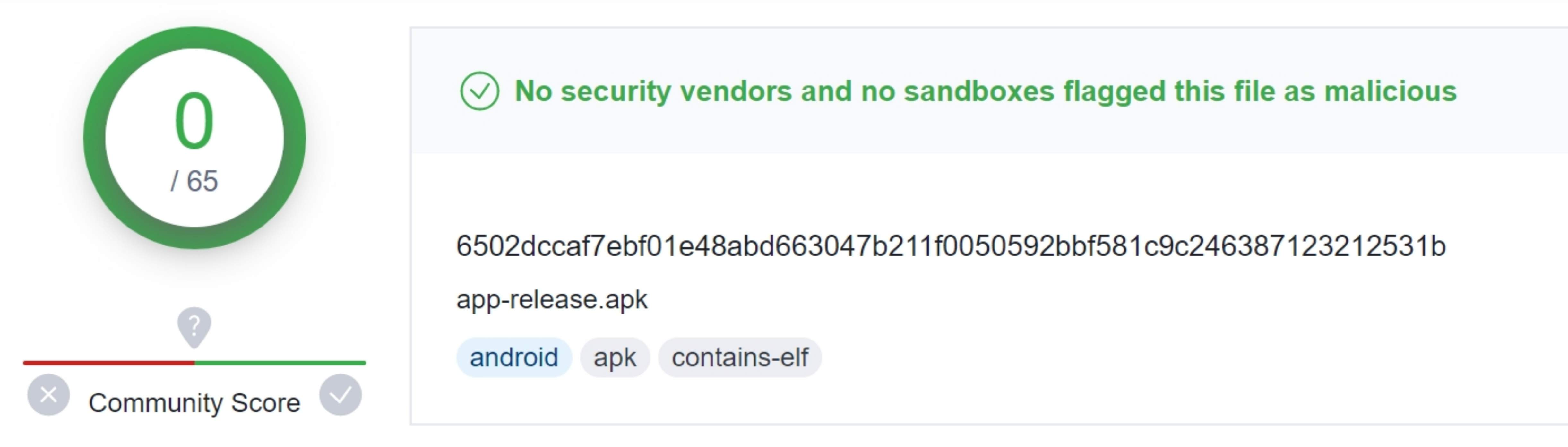}
\caption[]{Stealthiness evaluation with VirusTotal.}
\label{fig-stealthiness}
\end{figure}

We use VirusTotal~\cite{zhu2020benchmarking,zhu2020measuring},
an online static anti-malware scanning service widely used by researchers and industry practitioners,
to scan our attack application.
As shown in Fig.~\ref{fig-stealthiness},
our attack application can evade all 65 well-known anti-malware engines on VirusTotal,
including Avast, Avira, BitDefender, and other powerful engines.

\subsubsection{Run-Time Detection}
Since our malicious application can evade static detection,
we evaluate the stealthiness of our attack using dynamic malware detection.
Our evaluation shows that our attack can bypass built-in and popular third-party detection tools.

\begin{figure}[!htbp]
\centering
\includegraphics[width=\linewidth]{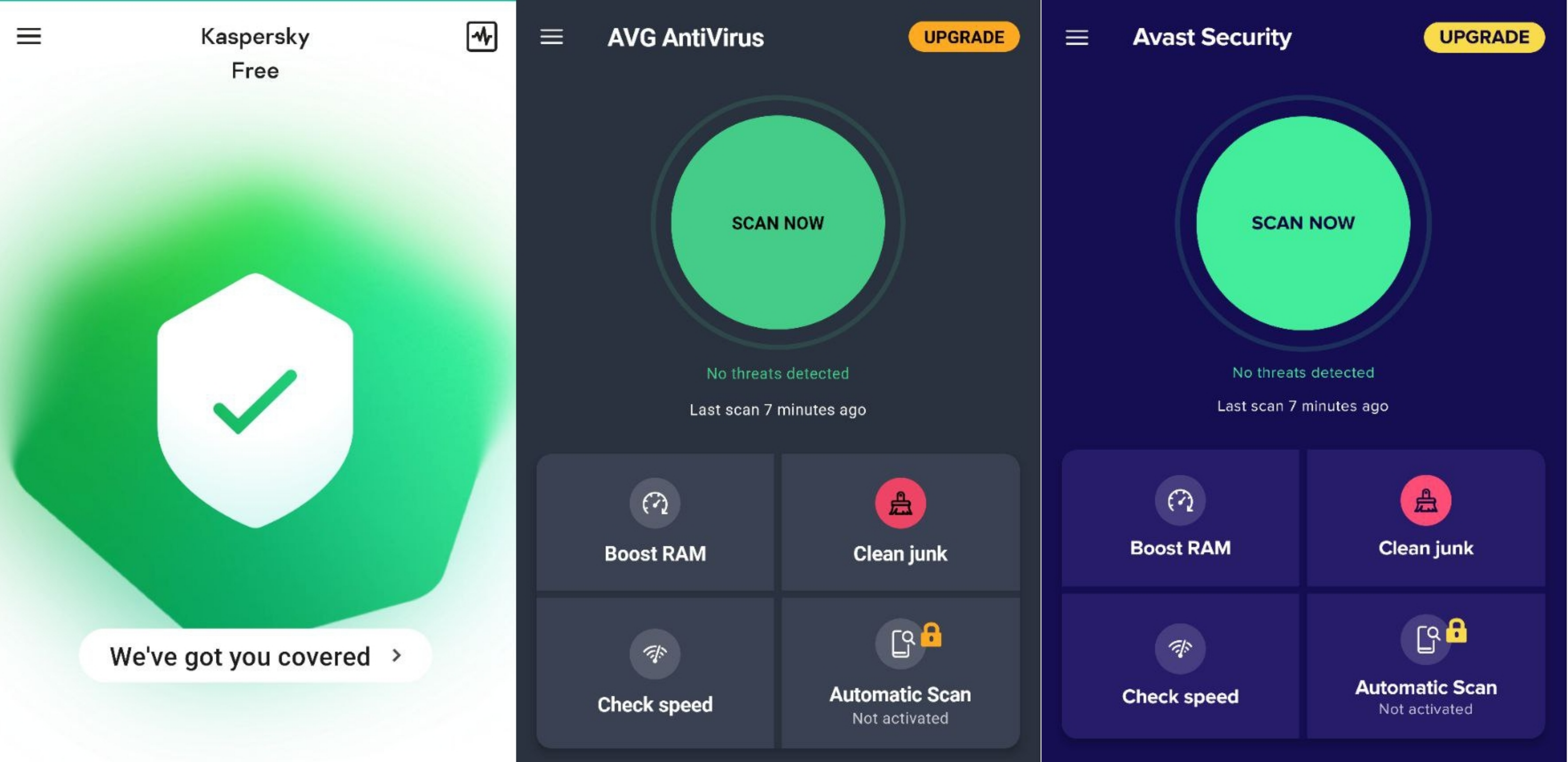}
\caption[]{Stealthiness evaluation with Avast, AVG, and Kaspersky.}
\label{fig-stealthiness2}
\end{figure}

\textbf{Built-in Detection.}
Most smartphones are only equipped with built-in detection engines provided by Android or their manufacturers.
We have successfully launched our attack on OPPO K10, Redmi K50, Xiaomi 9, and OnePlus 7Pro,
and the malicious application ran for several hours without being detected on these devices,
both of which show that our attack can be effective against widely used Android devices.

\textbf{Third-Party Detection.}
In addition, we also evaluate the stealthiness of our attack with popular third-party anti-malware engines.
We deploy and run our malicious application for several hours on OnePlus 7Pro,
which is equipped with Avast, AVG, and Kaspersky.
The result shown in Fig.~\ref{fig-stealthiness2} demonstrates that
our attack can evade all of these anti-malware engines,
highlighting the stealthiness of our attack.

\subsection{Evaluation in Real-World Settings}
We further evaluate the accuracy of our attack in real-world settings,
where users use the smartphone to perform random user behaviors.
In our experiment, 6 users use OPPO K10, Redmi K50, and Xiaomi 9 smartphones.
They then perform random fine-grained user behaviors
(17 behaviors in total described in Section~\ref{sec-setup}) for 2 minutes.

\begin{figure}[!htbp]
\centering
\includegraphics[width=2.7in]{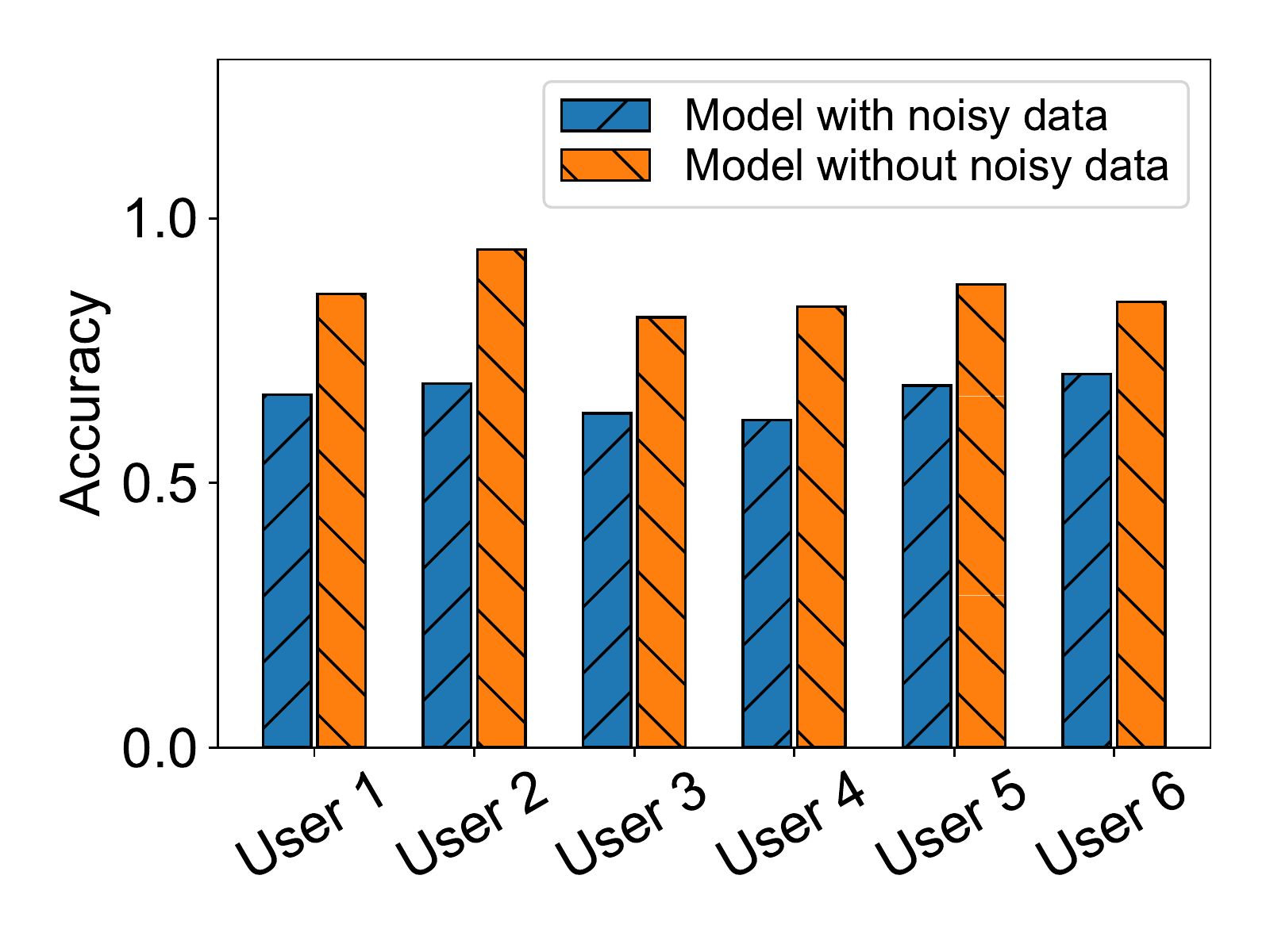}
\caption[]{Accuracy of our attack in real-world settings.}
\label{fig-realworld-accuracy}
\end{figure}

Fig.~\ref{fig-realworld-accuracy} shows that the average accuracy of
inferring user behaviors is 86\% for all four applications
when we use the model trained with noisy data
(66\% for the model trained without noisy data).
The difference in accuracy between these models has similarities to those
in the previous experiments shown in Fig.~\ref{fig-noise} and Fig.~\ref{fig-noise2}.
The accuracy is slightly reduced compared to the results in Section~\ref{sec-accuracy}
but still sufficient to ensure practical inferring with reasonable guesses based on the context of user behaviors.

\section{Mitigation and Future Work}
In this section, we discuss the possible mitigation approaches against the proposed attack and future work.

\subsection{Mitigation}
\textbf{Malware Detection.}
An intuitive mitigation approach is to rely on malware scanning or on-device malware detection
to detect abnormal behaviors in applications, such as frequent system call invocations.
However, as we have evaluated, neither the existing static malware detection
nor the runtime malware detection can detect the abnormal behaviors of our attack application.
On the one hand, since standard Linux system calls are invoked so frequently in normal OS operations,
invoking these system calls repeatedly would not be considered abnormal.
On the other hand, existing malware detection methods~\cite{qiu2020survey} focus more on
the permissions of the applications~\cite{arora2019permpair,zhu2023android}
rather than the invocation of system calls,
as well as behavior identification and classification~\cite{lei2019evedroid,guerra2022concept},
which is hard to integrate into Android systems.

\textbf{Obfuscation of Return Values.}
Another simple mitigation approach is to obfuscate the values of system calls.
Obfuscating the return values of system calls can confuse attackers,
which can be achieved by running random workloads in the background.
However, our attack can identify the state of the background music
and the PiP hover window as we have evaluated,
which means that these workloads are still observable.
Therefore, the main challenge is to determine the appropriate amount of these workloads.
On the one hand, excessive workloads would consume system resources and degrade system performance.
On the other hand, insufficient workloads would fail to mask user behaviors,
allowing attackers still can launch effective attacks.

\textbf{Access Control of Return Values.}
A more effective and practical approach is to use access control
to restrict the permissions of applications.
Our idea builds on the Android security model,
which uses discretionary access control (DAC) to isolate different applications.
Each application on Android can be considered a distinct user
since it is assigned a unique Linux UID and GID during installation,
and permissions are just strings associated with the application identifiers~\cite{backes2016demystifying}.

Role-based access control (RBAC) is a mechanism that allows users
to assign roles to applications and restrict the permissions of applications.
Consequently, a viable approach is to use RBAC to manage the permissions of each application,
i.e., to prevent applications with no permissions from invoking Linux system calls,
while allowing applications with permissions to invoke Linux system calls without restriction.

Android uses security-enhanced Linux (SELinux) to enforce mandatory access control (MAC) on all processes,
even those running with root/superuser privileges.
With this mechanism, all running applications are assigned roles.
Then their access to system resources can be monitored by the SELinux access manager
based on the security policy provided by the operating system.
Therefore, we could filter suspicious applications that invoke system calls
based on a whitelist by adding access rules to the security policy.

\subsection{Future Work}
Some directions for future work include classifying previously unseen user behaviors,
launching our attack on Apple devices, and implementing effective mitigations.

\textbf{Classifying Previously Unseen Behaviors.}
The first limitation of our approach is the use of supervised learning algorithms,
which allow the attacker to identify already-known user behaviors in training datasets.
However, it is not possible to identify classes of user behaviors
that were not trained during the training phase.
Therefore, future work may focus on identifying previously unseen applications and user behaviors,
which require semi-supervised or unsupervised learning rather than supervised learning.

\textbf{Launching Attack on Apple Devices.}
The second limitation of our approach is that, like most work,
we only focus on inferring user behaviors on Android systems.
Therefore, future work could focus more on the iOS mobile operating system
to fill the gap with inferring privacy on Apple smartphones with OS side-channel attacks.
And the iOS kernel is built on top of Mach and FreeBSD,
which is similar to the Linux kernel and has many system calls,
so it is possible to launch our attack on Apple devices.
And one of the main similarities between iOS and Android is that
the control of permissions is limited to access to contacts, access to storage, and other aspects.
Therefore, the attack against Android smartphones proposed in this paper
is likely to bypass the existing defense mechanisms for user privacy on iOS.

\textbf{Implementing Effective Mitigations.}
The third limitation of our approach is that
we only discuss the potential mitigations against the proposed attack,
but we never implement any of them to protect the privacy of user behaviors.
Thus, it would be interesting to develop effective malware detection techniques and
implement efficient but not excessive obfuscation or access control mechanisms for vulnerable system calls.

\section{Conclusion}
This paper proposes \textsl{EavesDroid}, an OS side-channel attack on Android smartphones that
enables unprivileged attackers to infer fine-grained user behaviors.
Without requiring additional privileges, \textsl{EavesDroid} permits the attacker to identify
fine-grained user behaviors with an accuracy rate of 98\% in 40ms
for the test set and 86\% in real-world settings.
And we show that our CNN-GRU classifier outperforms the DTW-KNN classifier by about 15\%,
while mix-max normalization and feature combination can improve the accuracy by about 6\% and 9\%, respectively.
In addition, \textsl{EavesDroid} works across a wide range of smartphones
running the same version of Android and can evade static and dynamic anti-malware detection,
highlighting the vulnerability of our attack and the need for countermeasures.
Therefore, we suggest performing effective malware detection,
obfuscating vulnerable return values with low overhead,
or restricting applications from accessing vulnerable return values
to enhance the security of Android smartphones.

\section*{Acknowledgements}
This work was supported by National Key R\&D Program of China under Grant No. 2022YFB3103800
and National Natural Science Foundation of China under Grant Nos. 61972295 and 62072247.

\bibliography{ref}
\bibliographystyle{IEEEtran}

\end{document}